\documentstyle[12pt,epsf]{article} 
\begin{document}
\newcommand{\dfrac}{\displaystyle \frac}
\renewcommand{\thefootnote}{\#\arabic{footnote}}
\newcommand{\ve}{\varepsilon}
\newcommand{\krig}[1]{\stackrel{\circ}{#1}}
\newcommand{\barr}[1]{\not\mathrel #1}
\newcommand{\beq}{\begin{equation}}
\newcommand{\eeq}{\end{equation}}
\newcommand{\Tr}{{\rm Tr}}

\thispagestyle{empty}

\hfill UPDATED VERSION, \today

\hfill hep-ph/9507418

\hfill CRN 95-26

\hfill TK 95 16

$\,$\vspace{0.4cm}

\begin{center}

{\large \bf   THE REACTION \boldmath{$\pi N \to \pi \pi N$} AT THRESHOLD}

\vspace{0.2cm}

{\large \bf IN CHIRAL PERTURBATION THEORY }

\vspace{1.1cm}
                              
{\large V. Bernard$^{\dag ,1}$, 
N. Kaiser$^{\S ,2}$, Ulf-G. Mei\ss ner$^{\ddag ,3}$}

\vspace{0.5cm}

$^{\dag}$Physique Th\'{e}orique, Centre de Recherches Nucl\'{e}aires \\ 
B.P. 20, F-67037 Strasbourg Cedex 2, France \\
Universit\'{e} Louis Pasteur de Strasbourg, Institut de Physique, \\
3--5 rue de l'Universit\'e, F-67084 Strasbourg, France 

\vspace{0.5cm}

$^{\S}$Technische  Universit\"{a}t M\"{u}nchen, Physik Department T39, \\ 
James-Franck-Stra\ss e, D-85747 Garching, Germany

\vspace{0.5cm}

$^{\ddag}$Universit\"at Bonn, Institut f\"ur Theoretische Kernphysik, \\
Nussallee 14-16, D--53115 Bonn, Germany

\vspace{0.5cm}
 
email: $^1$bernard@crnhp4.in2p3.fr, $^2$nkaiser@physik.tu-muenchen.de,
$^3$meissner@pythia.itkp.uni-bonn.de

\end{center}

\vspace{1.5cm}

\begin{center}

{\bf ABSTRACT:}

\end{center}

\vspace{0.1cm}

In the framework of heavy baryon chiral perturbation theory,
we give the chiral expansion for the $\pi N \to \pi \pi N$ threshold amplitudes
$D_1$ and $D_2$ to quadratic order in the pion mass.
The theoretical results  agree within one standard deviation with the
empirical values. We also derive
a relation between the two threshold amplitudes of 
the reaction $\pi N \to \pi \pi N$ and the $\pi \pi$ S--wave scattering 
lengths, $a_0^0$ and $a_0^2$, respectively, to order ${\cal O}(M_\pi^2)$.
We show that there are uncertainties mostly related to resonance excitation 
which make an accurate determination of the $\pi \pi$
scattering length $a_0^0$ from the $\pi \pi N$ threshold amplitudes at present
very difficult. The situation is different in the $\pi \pi$ isospin two
final state. Here, the chiral series converges and one finds $a_0^2 = -0.031
\pm 0.007$ consistent with the one--loop chiral perturbation theory prediction.


\vfill


\pagebreak

\vspace{0.5cm}

\section{Introduction}
\label{sec:int}

Elastic pion--pion scattering in the threshold region is the purest 
process to test our understanding of the chiral symmetry
breaking in QCD. Already in the early days of current algebra,  
Weinberg \cite{wein1} showed that the $\pi \pi$ S--wave scattering
lengths $a_0^I$ (with $I=0,2$ the total isospin of the two--pion system) 
vanish in the chiral limit of zero quark masses. In
particular, he predicted $a_0^0 = 7M_\pi^2 / (32 \pi F_\pi^2) = 0.16$ and 
$a_0^2 = (-2/7) \, a_0^0$. This prediction was further
sharpened by Gasser and Leutwyler \cite{gl83} \cite{gl84} in the 
framework of chiral perturbation theory (CHPT), 
which is the effective field theory
of the standard model at low energies. 
In Ref.\cite{gl84} a very accurate prediction 
for the isospin zero, S--wave scattering length
was given,  $a_0^0 = 0.20 \pm 0.01$, 
which amounts to a 25$\%$ increase compared 
to the current algebra value.  The main assumption
underlying this result is that the order parameter 
of the chiral symmetry breaking, 
$B =-<0|\bar q q|0>  / F_\pi^2$ (with 
$<0|\bar q q|0>$ the scalar quark condensate and  
$F_\pi \simeq 93$~MeV the pion decay constant)
 is considerably larger than
$F_\pi$, $B \gg F_\pi$, which follows e.g. from 
the standard analysis to determine the
 light quark mass ratios from the Goldstone
boson masses. Another scenario, 
in which the quark condensate is very much smaller 
and consequently $B \simeq F_\pi$ has been
discussed in Refs.\cite{stern} \cite{orsay}. 
In this approach, the ratio of the average 
u and d mass to the strange quark mass is 
decreased, typically $2 m_s /(m_u + m_d) < 10$ and 
the resulting scattering length 
$a_0^0$ increases, typically $a_0^0 \ge 0.27$. 
To settle this very important issue, it is mandatory to determine the 
$\pi \pi$ S--wave scattering lengths within an accuracy of
about 20$\%$ (or better). 
For a review on these topics, see e.g. \cite{ulfrev}. 

It is, however, not straightforward to determine the $\pi \pi$ phase 
shifts in the threshold region experimentally. A few possible 
candidates are $K_{\ell 4}$--decays, pionic molecules or the 
reaction $\pi N \to \pi \pi N$. It is this latter process we will
be concerned with in the following. To be more precise, consider single 
pion production in the threshold region and above. Already Weinberg 
\cite{wein2} pointed out that the one--pion exchange diagram  
contains the four--pion vertex (with one pion leg off--shell). 
This opens two possibilities of
extracting the on--shell $\pi \pi$ interaction. 
First, one can consider peripheral 
processes at higher energies but low momentum transfer 
and try to isolate the pion--pole by standard Chew--Low type techniques 
(for recent work in this direction, see e.g. \cite{russ}).
Here, we will concentrate on the second way, namely to directly relate 
the two independent $\pi \pi N$ threshold amplitudes to the
scattering lengths $a_0^0$ and $a_0^2$. This approach was pioneered 
by Olsson and Turner \cite{ot} and has been used ever
since in most analyses of the threshold $\pi N \to \pi \pi N $ data.  
However, the Olsson--Turner approach predates QCD, it 
is not applicable any more beyond tree level. 
In particular, in its original formulation 
a parameter $\xi$, which is a measure of the 
type of chiral symmetry breaking, is left free.
In QCD, this parameter $\xi$ is exactly zero. A critical discussion of these
topics can be found in \cite{obkm}. 

On the other hand, over  the last few years 
an impressive series of experiments have measured
the total cross section for the processes  $\pi N\to\pi\pi N$ quite close to
threshold~\cite{kernel,pocanic,kernel2,kernel3,lowe}. 
Extracted values for the $\pi \pi$ scattering 
lengths are based on the Olsson-Turner approach 
with $\xi \ne 0$ \cite{pocanic,burkhard}. Therefore,
it is necessary to work out a more precise relation 
between the threshold $\pi \pi N$ amplitudes and the
$\pi \pi$ S--wave scattering lengths beyond tree level.  
A first step in this direction was made in
Refs.\cite{bkmp} \cite{bkmrev} where an improved low--energy 
representation accounting for corrections 
of order $M_\pi$ to the tree level relations was formulated. 
This led to novel low--energy theorems,
which can be directly compared with the threshold data. 
Not unexpectedly, one finds a satisfactory
description for the channel $\pi^+ p \to \pi^+ \pi^+ n$ 
and some significant deviations for the
process  $\pi^- p \to \pi^0 \pi^0 n$, 
reflecting the relative weakness/strength of the pion--pion 
interaction in a state with isospin two/zero (in the S--wave). 
Here, we present the results of a calculation to one 
loop accuracy in heavy baryon chiral perturbation theory 
(HBCHPT) \cite{jm} \cite{bkkm}, which accounts
for {\it all} corrections up--to--and--including ${\cal O}(M_\pi^2)$ 
to the tree level result and which
is therefore sensitive to the one--loop corrections to 
the $\pi \pi $ scattering lengths, besides many
other contributions. 

The paper is organized as follows. In sections~\ref{sec:pre1},
\ref{sec:pre2} and \ref{sec:pre3}
we formulate the problem, give necessary kinematics and discuss briefly the
effective Lagrangian that will be used. Section~\ref{sec:loop} contains 
the principal
results of this paper, namely the Born, one loop and counterterm contributions
to the two threshold $\pi \pi N$ amplitudes up--to--and including 
order $M_\pi^2$.
Numerical results are discussed in section~\ref{sec:results} and 
a short summary is
given in section~\ref{sec:summary}. 
Some technicalities are relegated to the appendices.

\section{Prelude I: Threshold kinematics for \boldmath{$\pi N \to \pi \pi N$}}
\label{sec:pre1}
Consider the process 
\beq
\pi^a(k) + N(p_1) \to \pi^b(q_1) + \pi^c(q_2) + N(p_2) \, \, , 
\eeq
 with $'a,b,c'$ pion isospin indices. 
$N$ denotes the nucleon (neutron or proton).
At threshold and in the centre--of--mass frame,
 we have $q_1 = q_2 = ( M_\pi, 0,0,0)$, with $M_\pi$ the pion
mass. Using the pseudoscalar
quark density $P^a (x) = \bar{q}(x) i \gamma_5 \tau^a q(x)$ as 
interpolating pion field, standard LSZ reduction 
leads to
$$ - \int d^4 x d^4 y <N|T [ P^a(0) P^b (x) P^c (y) ] |N> 
{\rm e}^{i(q_1 \cdot x + q_2 \cdot y)}$$
\beq
 = \dfrac{G_\pi^3}
{(M_\pi^2 - q_1^2) \,(M_\pi^2 - q_2^2) \, (M_\pi^2 - k^2)} 
\,\, T^{\rm off-shell} \, \, ,
\label{lsz}
\eeq
with $G_\pi$ defined via
\beq
<0|P^a (0)| \pi^b> = \delta^{ab} \, G_\pi \quad.
\eeq
At threshold, the on--shell amplitude in the $\pi^a N$
centre--of--mass  system can be expressed in terms
of two threshold amplitudes,
\beq
T_{\rm cms}^{\rm on-shell} \equiv T \, 
= \, i \, {\vec \sigma} \cdot {\vec k} \left[ D_1 (\, \tau^b \delta^{ac} +
\, \tau^c \delta^{ab} ) \, + \, D_2 \, \tau^a \delta^{bc} \right] \, \, .
\label{defd}
\eeq
The quantities $D_{1,2}$ in eq.(\ref{defd}) are related to the commonly used
amplitudes ${\cal A}_{2I,I_{\pi \pi}}$, with $I$ the
total isospin of the initial $\pi N$ system 
and $I_{\pi \pi}$ the isospin of the two--pion system in the
final state, via
\begin{equation}
{\cal A}_{32} \, = \, \sqrt{10} \, D_1, \quad {\cal A}_{10} \, = \, -2D_1 \,
- \, 3D_2 \quad .
\label{adef}
\end{equation}
What we are after is the chiral expansion of the $D_1$ and $D_2$. It takes the
form (with ${\cal D}$ a generic symbol for $D_{1,2}$)
\beq
{\cal D} = f_0 + f_1 \, \mu + f_2 \, \mu^2 
+ \ldots, \quad \mu \equiv M_\pi / m \, \, ,
\label{dexp}
\eeq
modulo logarithms and we have introduced 
the pion to nucleon mass ($m$) ratio, $\mu$.

\section{Prelude II: Evolution and formulation of the problem}
\label{sec:pre2}
The modulus of the
threshold amplitude is determined by the extrapolation
of the measured  total cross section in the threshold region
via
\begin{equation}
| {\cal A}(\pi\pi N) |^2 = \lim_{T_\pi\to T_\pi^{\rm th}}
{ \sigma(\pi N\to\pi\pi N) \over C \, S \,(T_\pi-T_\pi^{\rm th})^2  }
\label{extrapol}   
                                               \end{equation}
where $T_\pi$ is the incident laboratory pion kinetic energy, $S$ is a Bose
symmetry factor ($S=1/2$ if the final two pions are identical, otherwise
it is unity), and
\begin{equation}
      C = M_\pi^2 \, \left(1\over128\pi^2\right) \sqrt3\, (2+\mu)^{1/2}\,
      (2+3\mu)^{1/2} \, (1+2\mu)^{-11/2} \, \, .
\end{equation}
The threshold modulus has been obtained in this way for the five charge states
initiated by $\pi^\pm p$~\cite{burkhard}.  Explicit isospin violation due to
the electromagnetic mass differences has been removed through the kinematics of
the threshold $T_\pi^{\rm th}$ value and the threshold amplitude
modulus is assumed to be isospin invariant.\footnote{A more systematic
  study of isospin violation is certainly needed.}
By Watson's theorem~\cite{watson} the threshold amplitude has the phase of the
initial elastic $J^P={1\over2}^+$ 
P--wave $\pi N$ scattering amplitude (up to an overall sign).  The
threshold production amplitude complex phase is then
$\delta_{31}\simeq-4^\circ$ for initial $\pi N$ isospin 3/2 and
$\delta_{11}\simeq2^\circ$
for initial isospin 1/2, as given by the respective phases evaluated at the
cms momentum of 213.6 MeV (the $\pi \pi N$ threshold).
  The threshold production amplitudes are thus nearly
real.  At threshold the final $\pi\pi$ S--wave 
state must have isospin 0 or 2 by
extended Bose symmetry and hence there are only two independent threshold
amplitudes as discussed before. From the
measured process amplitude moduli a unique value of ${\cal A}_{10}$
 and ${\cal A}_{32}$ can be found up to an overall sign. We note that
the sign is fixed by the chiral expansion (as discussed below).

In the Olsson--Turner approach \cite{ot} (with $\xi = 0$), it follows  that 
\begin{equation}
\def\arraystretch{1.5}
\begin{array}{rcl}
   {\cal A}_{32} &=& -2\sqrt{10} \pi \dfrac{g_{\pi N}}{m} \,
   \Bigl[ \dfrac{a^2_0}{M_\pi^2} + d_2 \Bigr] \\
   {\cal A}_{10} &=& \phantom+ 4 \pi \dfrac{g_{\pi N}}{m} \,
    \Bigl[\dfrac{a_0^0}{M_\pi^2} + d_0\Bigr]
\end{array}
\label{a_I}
\end{equation}
with $g_{\pi  N} =13.4$ the strong pion--nucleon coupling constant.
The above
result is a consequence of the dominance of the pion exchange and contact
diagrams. To lowest order, 
the two ``shift" constants $d_I$ arise from the
"sub-leading" diagrams involving three  pion absorptions/emissions
on the nucleon line, compare Fig.~1.
The $d_I$ are of order ${\cal O}(M_\pi)$. In the context of QCD, 
the relations Eq.(\ref{a_I}) are equivalent to 
the tree level CHPT results if the
$a_0^I$ are the tree level $\pi \pi$ scattering lengths {\`a} la Weinberg. The 
improved representation of \cite{bkmp} \cite{bkmrev}, 
which is based on the first
corrections to the Olsson--Turner result, takes the form
\begin{equation}
\def\arraystretch{1.5}
\begin{array}{rcl}
   {\cal A}_{32} &=& -2 \sqrt{10} \pi \, \dfrac{g_{\pi N}}{m} \, 
   \bigl( 1 + \frac{7}{2} \mu
   \bigr) \,   \Bigl[ \dfrac{a_0^2}{M_\pi^2} + \tilde{d}_2
                      M_\pi^2  \Bigr] \\
   {\cal A}_{10} &=& \phantom+ 4 \pi \, \dfrac{g_{\pi N}}{m} \,
   \bigl( 1 + \frac{37}{14} \mu
   \bigr) \Bigl[\dfrac{a_0^0}{M_\pi^2} +
                    \tilde{d}_0 M_\pi^2  \Bigr]
\end{array}
\label{a_Inew}
\end{equation}
where the new shift constants $\tilde{d}_{0,2}$ have the form
\beq
\tilde{d}_I = \tilde{d}_I^0  \, + \,
\tilde{d}_I^1 M_\pi  \, + \, \tilde{d}_I^2 M_\pi^2 \, +
\ldots \, , \quad I = 0 , 2
\label{shift}
\eeq
modulo logs. One notices that the correction of order $M_\pi$ is
comparable in size to the leading term (approximately
40$\%$ and 50$\%$ for ${\cal A}_{10}$ and ${\cal A}_{32}$, respectively).
Therefore, it is mandatory to
calculate (at least) the coefficients $\tilde{d}_I^0$. Also, at that order the
one--loop corrections to the S--wave $\pi \pi$ scattering lengths
appear \cite{gl83, gl84}. The problem investigated here is thus 
{\it nothing but the
calculation of the constants $\tilde{d}_I^0$ in eq.(\ref{shift})}.

\section{Prelude III: Effective Lagrangian}
\label{sec:pre3}
In this section, we will briefly discuss the chiral effective 
pion--nucleon Lagrangian as well as the pionic one
underlying our calculation. Many additional details are spelled out
in Refs.\cite{gl84} \cite{bkkm}.
To explore in a systematic fashion the consequences of spontaneous and explicit
chiral symmetry breaking of QCD, we make use of
 baryon chiral perturbation theory
(in the heavy mass formulation) \cite{jm} (HBCHPT). 
The nucleons are considered as
extremely heavy. This allows to decompose the nucleon Dirac spinor into "large"
$(H)$ and "small" ($h)$ components 
\beq
\Psi(x) = e^{-i m v \cdot x } \{ H(x) + h(x)\}  \, \, ,
\label{heavy}
\eeq
with $v_\mu$ the nucleon four-velocity, $v^2 = 1$, and the velocity eigenfields
are defined via $ \barr v H = H$ 
and $\barr v h = - h$.\footnote{The role of $v_\mu$ is to
single out a particular reference frame \cite{eckerrev}, 
here the $\pi^a N$ centre--of--mass frame.} 
Eliminating the "small" component field $h$ (which generates $1/m$
corrections), the leading order chiral $\pi N$ Lagrangian reads
\beq
{\cal L}_{\pi N}^{(1)} 
= {\bar H} ( i v\cdot D + \krig{g}_A S \cdot u ) H \, \, \, .
\label{lpin1}
\eeq 
Here the pions are collected in a SU(2) matrix-valued field $U(x)$
\beq 
U(x) = {1\over F} \biggl[ \sqrt{ F^2 - {\vec \pi} \, (x)^2 } + i \vec \tau
\cdot \vec \pi \, (x) \biggr] \label{gauge}
\eeq
with $F$ the pion decay constant in the chiral limit and the so-called
$\sigma$-model gauge has been chosen which is of particular convenience for our
calculations in the nucleon sector. In eq.(\ref{lpin1}) $D_\mu = \partial_\mu +
\Gamma_\mu$ denotes the nucleon chiral covariant derivative, $S_\mu$ is a
covariant generalization of the Pauli spin vector, $\krig{g}_A$ the
nucleon axial vector coupling constant  in the chiral limit,
$u_\mu = i u^\dagger \nabla_\mu U u^\dagger $,
with $u = \sqrt{U}$ and $\nabla_\mu$ the covariant derivative acting on the
pion fields. To leading order, ${\cal O}(q)$, one has to calculate tree 
diagrams from
\beq
 {\cal L}^{(1)}_{\pi N} + {F^2 \over 4}{\rm Tr}\bigl\{ \nabla^\mu U
\nabla_\mu U^\dagger + \chi_+ \bigr\}  \, \, \, ,
\quad \chi_+ = u^\dagger \chi u^\dagger + u \chi^\dagger u 
\label{leff1}
\eeq
where the second term is the lowest order mesonic chiral effective Lagrangian,
the non-linear $\sigma$-model coupled to external sources. 
The quantity $\chi$ contains the light quark mass $\hat m$ and
external scalar and pseudoscalar sources (the latter are actually needed to
compute correlators of the pseudoscalar quark density as in
eq.(\ref{lsz}) ) .  
Later, we will need the five--pion--nucleon vertex. Expanding $u$ in powers
of $\phi = \vec{\tau} \cdot \vec{\pi} / F$ gives
\beq
u = 1 + {i \over 2} \phi - {1 \over 8} \phi^2 + {i \over 16} \phi^3
- {5 \over 128} \phi^4 + {7i \over 256} \phi^5 + {\cal O}(\phi^6) \, \, ,
\label{uexp}
\eeq
and $u_\mu$ follows correspondingly, $u_\mu = i \lbrace u^\dagger , 
\partial_\mu u \rbrace$.    
  
To one--loop accuracy, i.e. order ${\cal O}(q^3)$, 
one has to consider tree graphs from
\beq
{\cal L}_{\rm eff} = 
{\cal L}_{\pi N}^{(1)} + {\cal L}_{\pi N}^{(2)} + {\cal L}_{\pi N}^{(3)} 
+ {\cal L}_{\pi \pi}^{(2)} + {\cal L}_{\pi \pi}^{(4)} \, \, .
\label{leff}
\eeq
where the structure of $ {\cal L}_{\pi N}^{(2)}$ 
is discussed in detail in \cite{bkmrev}
and, on a pedagogical level, in \cite{prag}. 
All terms in $ {\cal L}_{\pi N}^{(2)}$
are finite. The first divergences appear to ${\cal O}(q^3)$ in HBCHPT,
the corresponding determinant has been worked out by Ecker \cite{ecker},
\beq
{\cal L}_{\pi N}^{(3, {\rm div})} 
= \sum_{i=1}^{22} \, b_i \, \bar{H} \, O_i \, H \, \, , \quad 
b_i = b_i^r (\lambda) + \dfrac{\beta_i}{F^2} \, L \, \, 
\label{lpin3d}
\eeq
where the $O_i$ are monomials in the fields, 
$\lambda$ is the scale of dimensional regularization
and the $b_i$ differ by a factor $(4  \pi F_\pi )^2$ from the ones 
in \cite{ecker}, and
\beq
L = \dfrac{\lambda^{d-4}}{16 \pi^2} \biggl[ \dfrac{1}{d-4} - \dfrac{1}{2}
\bigl( \ln(4\pi ) - \gamma_E +1 \bigr) \biggr] \, \, ,
\label{defL}
\eeq
with $ \gamma_E = 0.5772$ the Euler--Mascheroni constant.
There are also
terms in ${\cal L}_{\pi N}^{(3)}$ with finite coefficients. The
corresponding low--energy constants will be estimated by resonance exchange.
It is important to note that some of the terms in ${\cal L}_{\pi N}^{(2,3)}$ 
are simply $1/m$ and $1/m^2$ corrections from the original Dirac
Lagrangian, like e.g. $\bar{H} D^2 / (2m) \, H$ (for details, see \cite{bkkm}).
${\cal L}_{\pi N}^{(2)}$ contains  terms proportional to the
low--energy constants $c_1,c_2,c_3,c_4$. The latter are related to the $\pi N$
$\sigma$-term and $\pi N$ scattering lengths as discussed below.  
In order to restore unitarity in a perturbative fashion, one has to include
(pion) loop diagrams. In HBCHPT, there exists a strict one-to-one
correspondence between the expansion of any observable in small external
momenta {\it and} quark masses and the 
expansion in the number of (pion) loops. In 
what follows we will work within the one-loop approximation corresponding to
chiral power ${\cal O}(q^3)$.
To obtain all contributions at order $q^3$ one
has to supplement  the chiral effective Lagrangian by the additional term 
${\cal L}^{(4)}_{\pi \pi}$ \cite{gl84}. It  serves to cancel some of
the divergences of certain loop diagrams and contains the
mesonic low--energy constants $\ell_{1,2,3,4}$. The latter encode information
about the chiral corrections to the $\pi \pi$ scattering lengths.
 We use the following form  of  ${\cal L}^{(4)}_{\pi \pi}$ \cite{gss},
$${\cal L}^{(4)}_{\pi \pi} = \dfrac{\ell_1}{4} 
\bigl(\Tr \nabla_\mu U \nabla^\mu U^\dagger \bigr)^2 +
\dfrac{\ell_2}{4} \Tr (\nabla_\mu U \nabla_\nu U^\dagger) 
\Tr (\nabla^\mu U \nabla^\nu U^\dagger)
+ \dfrac{\ell_3}{16} \bigl( \Tr \chi_+ )^2  $$
\beq
+ \dfrac{\ell_4}{16} \biggl\lbrace 
2\Tr ( \nabla_\mu U \nabla^\mu U^\dagger ) \Tr \chi_+ + 
2 \Tr(\chi^\dagger U \chi^\dagger U +
\chi U^\dagger \chi U^\dagger ) - 4 \Tr (\chi^\dagger \chi ) - 
(\Tr \chi_-)^2 \biggr\rbrace  
+ \ldots
\label{lpipi4}
\eeq
where the ellipsis stands for other terms of order $q^4$ which do, 
however, not 
contribute in our case. The finite pieces $\ell^r_i$ of the
low--energy constants
$\ell_i$ in eq.(\ref{lpipi4}) are renormalization
scale dependent and are related to the $\bar{\ell}_i$ of ref.\cite{gl84} via
$$ \bar{\ell}_1 = 96 \pi^2 \ell^r_1 (\lambda )
 - 2 \ln  \dfrac{M_\pi}{\lambda}  \, \, , \quad
\bar{\ell}_2 = 48 \pi^2 \ell^r_2 (\lambda )
- 2 \ln  \dfrac{M_\pi}{\lambda}  \, \, , $$
\beq
\bar{\ell}_3 = -64 \pi^2 \ell^r_3 (\lambda )  
- 2 \ln  \dfrac{M_\pi}{\lambda}  
\, \, , \quad
\bar{\ell}_4 = 16 \pi^2 \ell^r_4 (\lambda )
- 2 \ln  \dfrac{M_\pi}{\lambda} \, \, ,
\label{elli}
\eeq
and their actual values will be discussed later. From the Lagrangian
eq.(\ref{lpipi4}), one derives the chiral corrections to the S--wave
$\pi \pi$ scattering lengths \cite{gl84} (we do not exhibit the
explicit scale--dependence of the $\ell_i^r$ any more)
\beq
a_0^0 = {7M_\pi^2 \over 32 \pi F_\pi^2} \biggl[ 1 + \biggl({M_\pi \over 4 \pi
F_\pi} \biggr)^2 \biggl( {5\over 2} - 9 \ln {M_\pi \over \lambda} \biggr) + {2
M_\pi^2 \over 7F_\pi^2 } \biggl( 
20 \ell_1^r + 20 \ell_2^r +5 \ell_3^r  + 7 \ell_4^r \biggr) \biggr] 
\label{a004}
\eeq
\beq
a_0^2 = - {M_\pi^2 \over 16 \pi F_\pi^2} \biggl[ 1 + \biggl({M_\pi \over 4 \pi
F_\pi} \biggr)^2 \biggl( 3 \ln {M_\pi \over \lambda}-{1\over2}  \biggr) + {2
M_\pi^2 \over F_\pi^2 } \biggl( 
-4 \ell_1^r  -4 \ell_2^r - \ell_3^r  +  \ell_4^r  \biggr) \biggr] 
\label{a024}
\eeq
with $M_\pi$ and $F_\pi$ the empirical values (i.e. the corresponding
chiral corrections have been accounted for, see also \cite{gm}).
We have now assembled all tools
for calculating the one loop corrections for $\pi N \to \pi \pi N$.

\section{\boldmath{$\pi N \to \pi \pi N$} to one loop}
\label{sec:loop}
Before presenting the results of the calculation, 
some general remarks are in order. The
various contributions to the
chiral expansion of the invariant functions 
$D_{1,2}$ to order $q^3$ can be grouped as
\beq
{\cal D}^{(3)} = {\cal D}^{\rm Born} + {\cal D}^{\rm one-loop} 
+ {\cal D}^{\rm ct} \, \, ,
\eeq
where ${\cal D}^{\rm Born}$ 
subsumes the lowest order 
relativistic tree graphs and all kinematical corrections 
to it (which are suppressed by powers of $1/m$), 
${\cal D}^{\rm one-loop}$ the generic one loop
graphs and ${\cal D}^{\rm ct}$ the counter terms, 
which absorb the divergences from the loops and
there are, of course, additional finite ones. 
At threshold, the calculation simplifies since we have the selection rules
\beq
S \cdot q_1 = S \cdot q_2 = v \cdot (q_1 - q_2) = 0 \, \, ,
 \quad v \cdot q_1 = v \cdot q_2 =  M_\pi \, \, \quad
v \cdot k = 2M_\pi + {\cal O}(\frac{1}{m})  \, \, .
\label{selec}
\eeq
Also, from the start 
we will perform mass and coupling constant renormalization, 
as explained in the
context of the Born graphs (see below).  
We will give a fairly detailed description of the 
renormalization of the remaining divergences since 
that will serve as an excellent check on the
calculations. Furthermore, the extraction of the
 various low--energy constants and a thorough 
discussion of the related uncertainties is mandatory 
to really filter out the sensitivity of the
threshold $\pi \pi N$ amplitudes to the S--wave $\pi  \pi $ scattering lengths.

\subsection{Renormalized Born terms}
\label{sec:born}
{}From the three--pion--nucleon seagull and the 
pion--pole diagram,\footnote{ It is important to remark that this
splitting has no physical meaning. The contributions of the threshold
amplitudes $D_{1,2}$ coming from the seagull and the pion pole depend on the
choice of interpolating pion field, i.e. how one parametrizes $U(x)$ 
through some pion field.  The sum as a physical quantity 
is of course unique and independent of the particular choice.}    
one immediately finds the 
leading ${\cal O}(q)$ contribution to $D_{1,2}$
\beq
D_1 = \dfrac{\krig{g}_A}{8 F^3} \, \, , \quad 
D_2 = - \dfrac{3\krig{g}_A}{8 F^3} \, \, .
\label{dtree}
\eeq
It is then most economic to calculate the relativistic tree graphs 
and expand the
result in powers of $\mu$, see Fig.~1.
This gives automatically all kinematical corrections
to eq.(\ref{dtree}) and reads to order ${\cal O}(q^3)$
\beq
D_1^{\rm Born} 
= \dfrac{\krig{g}_A}{8 F^3} \biggl[ 1+ \dfrac{7}{2}\mu - \mu^2 \biggl(
\dfrac{1}{8}+\dfrac{3}{2}g_A^2 \biggr) \biggr] \, \, ,  
D_2^{\rm Born} 
= \dfrac{\krig{g}_A}{8 F^3} \biggl[ -3- \dfrac{17}{2}\mu + \mu^2 \biggl(
\dfrac{11}{8}+ 2 g_A^2 \biggr) \biggr] \, \, .  
\label{dborn}
\eeq
In what follows, we have to renormalize the factor $\krig{g}_A /F^3$
(i.e. the chiral limit value) to the physical value $ g_{\pi N}
/ (m F_\pi^2)$. This renormalization procedure subsumes a host
of loop and counter term corrections. For doing that,
we first have to consider the pion mass, decay constant and so on \cite{gl84}. 
To one loop 
we  have for the pseudoscalar coupling 
$G_\pi$ (cf. Fig.~2)
\beq 
G_\pi = G \biggl[ 1 + \dfrac{M^2}{F^2} \biggl( 2 \ell_3^r  +
\ell_4^r  - \dfrac{1}{16 \pi^2} \ln  \dfrac{M}{\lambda}
 \biggr) \biggr] \, \, ,
\eeq
with $G$ the chiral limit value of $G_\pi$. Similarly, we find
for the pion decay constant
\beq 
F_\pi = F \biggl[ 1 + \dfrac{M^2}{F^2} \biggl(
\ell_4^r  - \dfrac{1}{8 \pi^2} \ln  \dfrac{M}{\lambda}
 \biggr) \biggr] \, \, ,
\eeq
and the pion mass renormalization reads
\beq 
M_\pi^2 = M^2 \biggl[ 1 + \dfrac{M^2}{F^2} \biggl(
2 \ell_3^r  + \dfrac{1}{16 \pi^2} \ln  \dfrac{M}{\lambda}
 \biggr) \biggr] \, \, ,
\eeq
with $M^2 = 2 \hat{m} B$ the leading term in the quark mass expansion of
the pion mass squared. With that, the pion propagator takes the form
(in the $\sigma$--model gauge), 
\beq 
\dfrac {i \, Z_\pi}{q^2 - M^2_\pi } \, \, , \, 
Z_\pi = 1 - \dfrac{M^2}{F^2} \biggl[2L + 2 \ell_4 + \dfrac{1}{8 \pi^2}
\ln  \dfrac{M}{\lambda}  \, \biggr] \, \, .
\eeq
The pertinent diagrams contributing to the 
renormalization of the pion--nucleon vertex are
shown in Fig.~3. The appropriate one--loop graphs 
for $\pi N \to \pi \pi N$ which account for 
this renormalization will have to be identified and subtracted
accordingly,  see the next section. The coupling constant
renormalization can be written as
\beq
\dfrac{g_{\pi N}}{m} = \dfrac{g_A}{F_\pi} \biggl( 1 - 
\dfrac{2 b_{11} M_\pi^2}{F_\pi^2} \biggr) \, \, \, ,
\label{gtr}
\eeq
where the constant $b_{11}$ is finite. Its value is fixed from the
known Goldberger--Treiman discrepancy. 
However, there remains a finite contribution to $D_{1,2}$ from
the corresponding chiral power three Lagrangian,
\beq
{\cal L}_{\pi N}^{(3)} = b_{11} \dfrac{g_A}{F^2} \, \bar{H} \, i S \cdot
(D \chi_-) \, H \quad ,
\eeq
after
the $g_{\pi N}$ renormalization, eq.(\ref{gtr}), has been performed,
\beq
D_1^{\rm GTR} = 0 \, \, , \quad
D_2^{\rm GTR} = - g_A \, b_{11} \, \dfrac{M_\pi^2}{F_\pi^5} \, \, .
\label{d12gtr}
\eeq
This has to be accounted for. 

\subsection{One loop graphs}
\label{sec:1loop}
In Fig.~4, we show the 36 different one--loop diagrams 
that contribute at threshold (we
do not display graphs in which the two out--going pions are
interchanged, 
$b \leftrightarrow
c$). We have made use of the selection rules, 
eq.(\ref{selec}), and omitted all those diagrams
which according to these rule are equal to zero. 
This means that if one wants 
to extend this calculation
to kinematics above threshold, 
one has to consider many more diagrams than shown in
Fig.~4 since then the selection rules do not apply anymore.  
Of course, many of the graphs shown contribute to 
mass and coupling constant
renormalization. Concerning the chiral 
corrections to the pion-pion interaction, the
interesting diagrams are the ones numbered 16, 17 and 18. 
The calculation of all
these diagrams is straightforward but somewhat tedious. 
It is most economically done
in the basis of the loop functions defined in appendix~B 
of ref.\cite{bkmrev}. In 
appendix~A, we assemble some novel loop functions 
not considered in \cite{bkmrev}.

Putting all pieces together, one has 
(after renormalization of $F_\pi$, $G_\pi$ and $g_{\pi N}$),
$$D_1^{\rm loop} = \dfrac{g_A M_\pi^2}{32 \pi^2 F_\pi^5} \biggl[ 
\biggl(g_A^2 - \dfrac{1}{4} \biggr)
\ln \dfrac{M_\pi}{\lambda}  - \dfrac{25}{24} + \dfrac{g_A^2}{6} + \bigl(
\dfrac{g_A^2}{2} - \dfrac{7}{8} \bigr) \sqrt{3} \ln (2 + \sqrt{3}) - i
\dfrac{\pi}{4} \sqrt{3} g_A^2 + 10\, I \biggr] $$
\beq
+ \dfrac{g_A M_\pi^2}{F_\pi^5} \biggl[ 
\dfrac{g_A^2}{2} - \dfrac{1}{8} \biggr] \, L  \, \, , 
\label{d1l}
\eeq
$$D_2^{\rm loop} = \dfrac{g_A M_\pi^2}{32 \pi^2 F_\pi^5} 
\biggl[ \bigl( \dfrac{43}{4} -10 g_A^2 \bigr) \ln 
\dfrac{M_\pi}{\lambda}  - \dfrac{73}{24} + \dfrac{16 g_A^2}{3} - \bigl(
5 g_A^2 - \dfrac{1}{2} \bigr) \sqrt{3} \ln (2 + \sqrt{3}) + i
\dfrac{5 \pi}{2} \sqrt{3} g_A^2 + 4 \, I \biggr] $$
\beq
+ \dfrac{g_A M_\pi^2}{F_\pi^5} 
\biggl[ \dfrac{43}{8} - 5 g_A^2  \biggr] \, L \, \, , 
\label{d2l}
\eeq
with
\beq
I = \int_0^1 dx \dfrac{x}{\sqrt{(1-x)(1+3x)}} \arctan
\dfrac{x}{\sqrt{(1-x)(1+3x)}} = 0.6456 \, \, \, .
\label{idef}
\eeq
The imaginary part in eqs.(\ref{d1l},\ref{d2l}) is due to the diagrams numbered
$20, 21, 24, \dots , 28$. For these pion--nucleon rescattering type of graphs,
the pertinent loop functions have to be evaluated at $\omega = 2 M_\pi$,
which is well above the branch point $\omega_0 = M_\pi$. This 
is similar to the effect observed in the 
calculation of the threshold amplitudes
for the reaction $\gamma N \to \pi \pi N$, where one also finds an imaginary
part at threshold \cite{bkms}. 
If one calculates from eqs.(\ref{d1l},\ref{d2l}) 
the imaginary parts of the isospin amplitudes
${\cal A}_{10}$ and ${\cal A}_{32}$ and compares 
them to those demanded by Watson's
theorem   (using the experimental fit values for the real parts),
one makes the following observation.
The phase is approximately correct for the isospin 3/2 case but an 
order of magnitude too large with the wrong sign in the isospin 1/2 case. The
reason for this is that the tree level phases which we encounter here satisfy
$\delta_{11} = 4 \delta_{31}$ which does not hold for the empirical $P_{2I,1}$
phase at the $\pi\pi N$ threshold. Nevertheless, the appearance of the
tree level $\pi N$ phases in eqs.(\ref{d1l},\ref{d2l}) 
serves as a good check on the one-loop calculation. 
We encounter here the standard problem  in chiral perturbation theory
that for getting the imaginary parts
better, one has to perform a higher order calculation.
As discussed in the introduction,
$D_{1,2}$ have to be almost real, therefore we
will neglect in the following their imaginary parts. 
Next, we have to perform the remaining renormalizations to get rid of the terms
proportional to $L$.

\subsection{Renormalization}
\label{sec:renorm}
We proceed in two steps. First, we consider the divergences related to 
the pion--nucleon Lagrangian $ {\cal L}_{\pi N}^{(3)}$, eq.(\ref{lpin3d}).
The following operators as defined in Ref.\cite{ecker} give a non--vanishing 
contribution at threshold; $O_4, O_5, O_6 , O_7, O_9, O_{17}, O_{18}$ and 
$O_{20}$. Some of these are equivalent at threshold, these are $O_4$ and
$O_6$, $O_5$ and $O_7$ and the combination of $O_{17}$ plus $O_{18}$. The
corresponding $\beta$'s are $\beta_5 + \beta_7 = g_A (1 - g_A^2) / 2$,
$\beta_4 + \beta_6 = -g_A  / 2$ and $\beta_{17} + \beta_{18}
 =  (2 - 3 g_A^2) / 4$. The operator $O_9 = S \cdot u \Tr(\chi_+)$ 
has two terms with three pion
fields. First, there is a three--pion vertex from $u_\mu$. This 
contribution  is, however, completely contained in the renormalization
of $g_{\pi N}$. Second, there is a term with one pion from $u_\mu$
and two from $\Tr(\chi_+)$ with $\beta_9 = g_A (4 - g_A^2) / 8$.
Finally, there is the operator $O_{20} = i v \cdot D \Tr(\chi_+) 
+ {\rm h.c.}$. The relevant contribution has a vertex with
two pions coming from $\Tr(\chi_+)$ with
$\beta_{20} = - 9 g_A^2 /16$. Adding 
up all these counterterm countributions which in the case of 
$O_{17}$, $O_{18}$ and $O_{20}$ come from two step processes 
with a $\sigma \cdot k$ interaction  for the incoming pion $\pi^a$, 
we have (the scale--dependence of the $b_i^r$ is not made explicit)
\beq
D_1^{({\rm ct},3)} = \dfrac{M_\pi^2}{F_\pi^3} \biggl[ b_5^r  +
b_7^r  + \dfrac{g_A}{2 F_\pi^2} (1 - g_A^2 ) \, L \biggr] \, \, ,
\label{d1ct3}
\eeq 
\beq
D_2^{({\rm ct},3)} = \dfrac{M_\pi^2}{F_\pi^3} \biggl[ 2(b_4^r  +
b_6^r  + b_9^r  ) - 4 g_A ( b_{17}^r  +
b_{18}^r  + b_{20}^r  )
+ \dfrac{g_A}{ F_\pi^2} (5 g_A^2 -2 ) \, L \biggr] \, \, ,
\label{d2ct3}
\eeq 
where the terms $\sim g_A^3 \, L$ are cancelled by the infinities in 
$D_1^{\rm loop}$ and $D_2^{\rm loop}$, eqs.(\ref{d1l},\ref{d2l}),
respectively.  So we are left with the following divergent pieces:
\beq
D_1^{\rm div,loop+ct3} 
= \dfrac{g_A M_\pi^2}{F_\pi^5} \bigl(\dfrac{3}{8} \, L \bigr)
\, \, , \quad
D_2^{\rm div,loop+ct3}  
= \dfrac{g_A M_\pi^2}{F_\pi^5} \bigl(\dfrac{27}{8} \, L \bigr)
\, \, , \, 
\label{divrest}
\eeq
which have to be cancelled by the counter terms from 
$ {\cal L}_{\pi \pi}^{(4)}$. After renormaliztion of $F_\pi$, $G_\pi$
and $g_{\pi N}$,  the total contribution from $ {\cal L}_{\pi \pi}^{(4)}$
reads
\beq
D_1^{({\rm ct},4)} = \dfrac{g_A M_\pi^2}{F_\pi^5} \biggl[ -\dfrac{3}{8}
\, L - \dfrac{3}{2} \, \ell_2^r - \dfrac{1}{4} \, \ell_3^r +
 \dfrac{1}{4} \, \ell_4^r \biggr] \, \, ,
\label{d1div4}
\eeq
\beq
D_2^{({\rm ct},4)} = \dfrac{g_A M_\pi^2}{F_\pi^5} \biggl[ -\dfrac{27}{8}
\, L - 3 \, \ell_1^r  - \dfrac{1}{4} \, \ell_3^r  -  \dfrac{5}{4} \,
\ell_4^r \biggr] \, \, .
\label{d2div4}
\eeq
We remark that the low--energy constant $\ell_3$ does only appear
via the renormalization of the pion mass with the appropriate
insertion from ${\cal L}_{\pi\pi}^{(4)}$ for the pion hooking on 
to the nucleon (cf. graph 12 in Fig.4) since
$[\Tr(\chi_+)]^2$ has no four--pion vertex in the $\sigma$--model 
gauge.\footnote{We have checked that in other parametrizations 
of $U$ where $[\Tr(\chi_+)]^2$ has a four--pion vertex,
the final result is the same.}
Comparison of eqs.(\ref{d1div4},\ref{d2div4}) with eq.(\ref{divrest})
leads to the desired result, namely
\beq
D_i^{\rm div} = D_i^{{\rm div, loop}} + D_i^{{\rm div}, 3} +
D_i^{{\rm div}, 4} = 0 \, \, , \quad i = 1,2 \, \, .
\eeq
This cancellation of divergences serves as an important check on our 
calculation. The finite counter term contribution can be compactly written
as
\beq
D_1^{\rm ct, fin} = \dfrac{ M_\pi^2}{F_\pi^3} \biggl[ \dfrac{g_A}{F_\pi^2}
\, \biggl( \dfrac{\ell_4^r}{4} - \dfrac{\ell_3^r}{4} - 
\dfrac{3}{2}\ell_2^r \biggr) + \delta_1^r \,
  \biggr] \, \, ,
\label{d1fin}
\eeq
\beq
D_2^{\rm ct, fin} = \dfrac{ M_\pi^2}{F_\pi^3} \biggl[ -\dfrac{g_A}{F_\pi^2}
\, \biggl( 3 \, \ell_1^r + \dfrac{5}{4} \ell_4^r 
+ \dfrac{1}{4} \, \ell_3^r \biggr) + \delta_2^r \,
 \biggr] \, \, ,
\label{d2fin}
\eeq
where the $\delta_{1,2}^r$ subsume the appropriate contributions of the
$b_i^r$ plus additional finite pieces from $ {\cal L}_{\pi N}^{(3)}$ and
from $1/m$ suppressed corrections from $ {\cal L}_{\pi N}^{(2)}$. 
The estimation of these finite pieces will be discussed in the next section.

\subsection{Finite contributions and estimation of low--energy constants}
\label{sec:lecs}
The most difficult task is to pin down the finite terms $\delta_{1,2}^r$
in eqs.(\ref{d1fin},\ref{d2fin}). These can be split into two distinct 
contributions. First, there are $1/m$ suppressed terms with insertions
from the relativistic chiral order two Lagrangian ${\cal L}_{\pi N}^{(2)}$.
In Ref.\cite{bkmp} we had already shown that such terms cancel at order
$M_\pi^2$ which allowed to formulate low--energy theorems for $D_{1,2}$
independent of the corresponding low--energy
constants $c_i$. Here, we are working one order further
and thus such contributions appear, some examples are shown in Fig.5.
These are operators of dimension three which contribute at threshold. 
The other type of terms are related
to the values of the various $b_i^r (\lambda)$ due to the renormalization
and additional finite ones from $ {\cal L}_{\pi N}^{(3)}$. In the absence
of a complete data set to fit these, we 
will make use of the resonance saturation principle. 
This procedure will induce some uncertainty in our final results, see the
discussions in \cite{bkmrev} and \cite{ulfmit}.
The exception to this is the term $\sim b_{11}$ discussed in
section~\ref{sec:born}. This particular contribution can not be
explained by resonance exchange.

Consider first the contributions due to insertions from 
$ {\cal L}_{\pi N}^{(2)}$. These can be either calculated by performing
the path integral as in ref.\cite{bkkm} and expanding to the desired
order, or, more economically, by using the relativistic Lagrangian
$ {\cal L}_{\pi N}^{(2, {\rm rel})}$. It is this latter method we
are using.\footnote{In appendix \ref{appc}, we derive the result for
$D_{1,2}$ directly from the path--integral formulation of the
heavy--nucleon Lagrangian.}
 It is same method used in the calculation of the $1/m$ suppressed
(kinematical) corrections to the tree graphs, compare section~\ref{sec:born}.
It is based on the observation that all operators  which have 
fixed coefficients
like $1/m$, $1/m^2$, $g_A/m$, and so on are nothing 
but the expansion coefficients
of the relativistic theory (see e.g. Refs.\cite{bkkm},\cite{prag}).
The corresponding dimension two effective relativistic pion--nucleon Lagrangian
reads \cite{gss} \cite{krause}
$$
{\cal L}_{\pi N}^{(2, {\rm rel})} = {c}_1 \bar{\Psi} 
\Psi \Tr(\chi_+) + \dfrac{{c}'_2}{4m} \bar{\Psi}
i \gamma_\mu \stackrel{\leftrightarrow}{D}_\nu \Psi \Tr(u^\mu u^\nu ) -
\dfrac{{c}''_2}{8m^2} \bar{\Psi} 
\stackrel{\leftrightarrow}{D}_\mu \stackrel{\leftrightarrow}{D}_\nu \Psi
\Tr(u^\mu u^\nu ) $$
\beq
+ {c}_3 \bar{\Psi} u_\mu u^\mu \Psi
+ {c}_4 \dfrac{i}{4} \bar{\Psi} \sigma_{\mu \nu} [u^\mu, u^\nu ]\Psi
+ \ldots 
\label{lpinrel}
\eeq
where $\Psi$ denotes the relativistic nucleon field and 
the ellipsis stands for other terms not needed here. The ${c}_i$
are normalized such that we can identifiy them with the 
corresponding low--energy constants  of the heavy baryon
Lagrangian (truncated at order $q^2$) 
(for definitions, see e.g. \cite{bkmrev}).
Note that the constants $c_2'$ and $c_2 ''$ are related to the $c_4$
and $c_5$ of App. A in \cite{bkkm} and that they have been renamed in 
comparison to Ref.\cite{gss}. To leading order in the heavy baryon
Lagrangian, we have $c_2 = c_2' + c_2''$ in the notation of Ref.\cite{bkmrev}.
The contribution of these terms to the $D_{1,2}$ follows as,
\beq
D_1^{(c_i)} = \dfrac{g_A M_\pi^2}{m F_\pi^3} \biggl( -2c_1 + \dfrac{3}{2}
c_2' + 2c_2'' + 2 c_3 +\dfrac{1}{2} c_4 \biggr) \, \, ,
\label{d1ci}
\eeq
\beq
D_2^{(c_i)} =  - \dfrac{g_A M_\pi^2}{m F_\pi^3} \biggl(
c_2' + 2c_2'' +  c_4 \biggr) \, \, .
\label{d2ci}
\eeq
For the numerical evaluation of the $c_i$ and to the accuracy we are
working, we can fix them to order $q^2$ 
(i.e. just calculating tree graphs) from available
pion--nucleon scattering data. Therefore, the values presented here
will differ from previous estimates which included $q^3$
contributions. This difference is, of course, just one of the many
corrections of ${\cal O}(M_\pi^3)$ in $D_1$ and $D_2$ and only becomes
important if one wants to extend the calculation presented here to the
next order. 
First, we invoke the subthreshold expansion of the standard invariant
pion--nucleon amplitudes with the pseudovector Born terms subtracted 
(as indicated by the 'bar') \cite{lbII}
\beq
\bar{X} = \sum_{m,n} \, x_{m,n} \, \nu^{2m+k} \, t^n \, \, ,
X = \{A^+ , B^+, A^-, B^- \} \, \, , 
\label{xbar}
\eeq
with $t$ the invariant momentum transfer squared, 
$\nu = (s-u)/4m$ ($s,t,u$ are 
the conventional Mandelstam variables subject to 
the constraint $s+t+u= 2M_\pi^2
+2m^2$) and $k=1 \, (0)$ if the function considered is odd (even) 
in $\nu$. Retaining
terms to order ${\cal O}(\nu^2,t)$, one finds from eq.(\ref{lpinrel})
$$
\bar{A}^+ = -\dfrac{4 c_1 M_\pi^2}{F_\pi^2} + \dfrac{c_3}{F_\pi^2}(2M_\pi^2
-t) + \dfrac{2 c_2'' }{F_\pi^2} \nu^2 \, \, \, \, , 
\bar{B}^+ = \dfrac{2 c_2'}{F_\pi^2} \, \nu \, \, \, \, ,  $$
\beq
\bar{A}^- = -\dfrac{2 m  c_4 }{F_\pi^2} \, \nu \, \, \, \, ,
\bar{B}^- = \dfrac{1}{2F_\pi^2} + \dfrac{2 c_4 m}{F_\pi^2} \, \, \, ,
\label{subthr}
\eeq
where the first term in $\bar{B}^-$ stems from 
the celebrated Weinberg--Tomozawa
term \cite{wein1} \cite{tomo}. From that, we deduce the following relations
to order $q^2$,
$$
a_{00}^+ = \dfrac{2M_\pi^2}{F_\pi^2} (c_3 - 2c_1 ) \, \, \, \, , 
a_{00}^- = -\dfrac{2 m c_4}{F_\pi^2}  \, \, \, \, , 
a_{10}^+ = \dfrac{2  c_2''}{F_\pi^2}  \, \, \, \, , 
a_{01}^+ = -\dfrac{  c_3 }{F_\pi^2}  \, \, \, \, , 
$$
\beq
b_{00}^+ = \dfrac{2 c_2' }{F_\pi^2}  \, , \quad 
b_{00}^- = \dfrac{1}{2F_\pi^2} \, (1 + 4 c_4 m ) \, \, .
\label{abc}
\eeq
Further information is obtained for $c_1$ from the $\pi N$--$\sigma$ term
\cite{bkkm} \cite{gss}, $\sigma_{\pi N} (0) = - 4 c_1 M_\pi^2$ (to order
$q^2$) and for $c_3$ and
$c_4$ from the wave scattering volumina $c_0$ and $d_1$,
\beq 
c_0 = - \dfrac{c_3 }{2 \pi F_\pi^2} \, , \quad
d_1 = - \dfrac{1}{4 \pi F_\pi^2} \biggl( c_4 + \dfrac{1}{4m} \biggr)
\, \, . \label{scavol}
\eeq
 
We are left with terms related to ${\cal L}_{\pi N}^{(3)}$. There is the
finite contribution form the pion--nucleon vertex renormalization as 
discussed in section~\ref{sec:born}. It leads to $D_{1,2}^{\rm GTR}$
as given in eq.(\ref{d12gtr}).
The remaining contributions will be estimated by resonance exchange. This
works well in the meson sector \cite{reso}, it is, however, more complicated
in the baryon case since one can have excited nucleon intermediate states as
well as scalar and vector meson couplings to two or three pions. 
For our case, we estimate the genuine
counter terms from  $ {\cal L}_{\pi N}^{(3)}$ via single and double $N^\star$ 
excitations and  meson exchange, see Fig.~6. Notice that
at threshold, the vector meson contribution analogous to the scalar
meson one vanishes. To be precise, the $\rho$ 
has a P--wave coupling and the $3\pi \,
\omega$ vertex vanishes at threshold. In principle, there could also be
a contribution due to the $a_1 \, 3\pi$ coupling. However, the
branching ratio $a_1 \to (\pi \pi)_S \pi$ is so small that we can
safely neglect this contribution. We now turn to the non--vanishing parts. 
 First, we consider the $\Delta (1232)$. If one treats
it non--relativistically, all graphs with double--$\Delta$ excitations vanish
at threshold, as it is e.g. the case in the model of ref.\cite{oset}. These
diagrams are obviously of order $q^3$ and proportional to $\vec{q
  \,}_i$, which vanishes at threshold. Relativistically,
the double--$\Delta$ graphs first contribute at order $M_\pi^3$ to
$D_{1,2}$.

Second, there
is the Roper, $N^\star (1440)$. It has a very large width for decaying  into
a nucleon and two pions. The corresponding Lagrangian
${\cal L}_{N^\star N \pi \pi}$ for S--wave emission is
discussed in detail in appendix \ref{appb} since in the available
literature it is not treated in its most generality (as needed here).
On the other hand, ${\cal L}_{N^\star N \pi}$ is standard,
\beq
{\cal L}_{N^\star N \pi} = \dfrac{ g_A}{4}  \sqrt{R} \, \bar{\Psi}_{N^\star} 
\, \gamma_\mu \gamma_5 \, u^\mu  \, \Psi_N + {\rm h.c.}  \, \, ,
\label{rop1}
\eeq 
with $\sqrt{R} = 0.53 \pm 0.04$ from the total width
\cite{hoehrop} (calculated relativistically).\footnote{We use here the
  width as obtained from the speed plot, not the model--dependent
  Breit-Wigner fits, $\Gamma_{\rm tot} = 160 \pm 40$ MeV \cite{hoehrop} and as
  branching ratios BR($N^\star \to N \pi) =0.68$ and BR($N^\star \to
  N (\pi \pi)_S) = 0.075 \pm 0.025$ \cite{pdg}.}
Putting pieces together, the Roper contribution is
\beq
D_1^{N^\star} = 0 \, \, \,, \, \,
D_2^{N^\star} =  \dfrac{ (c_1^\star + c_2^\star )
\, g_A \, \sqrt{R} \, M_\pi^2}
{F_\pi^3 \,( m^\star - m ) } \, \, .
\label{d12roper}
\eeq
The coupling constant combination $ c_1^\star + c_2^\star  =
-1.56 \pm 3.35$~GeV$^{-1}$ follows from the partial width 
$\Gamma ( N^\star \to N (\pi \pi)_S )$ 
with both pions in an S--state \cite{pdg}. The details can be found in
appendix \ref{appb}.

The next resonance which decays into a nucleon and two pions is the 
$D_{13} (1520)$. Non--relativistically, this D--wave state cannot be
excited by the initial P--wave $\pi N$ system and thus we neglect such
a  possible contribution. Of course, at energies above threshold such
resonances become important, as witnessed e.g. by the study of the
reaction $\gamma p \to \pi^+ \pi^- p$ in \cite{oset2}. We remark that
our treatment of the baryon resonance contributions is in good agreement
with the partial wave--analysis of $\pi N \to \pi \pi N$ by
Manley et al. \cite{manl}. In their table VI one sees that at  the
lowest energy considered, only the Roper leads to  a sizeable cross
section. There is no sign of the $\Delta$ and the $D_{13}$
contribution is approximately one order of magnitude smaller than the
one from the $P_{11}$. This lends further credit to our argument of 
exclusively keeping the Roper excitation to estimate the baryon
resonance contribution to the threshold $\pi \pi N$ amplitudes.  
Third, there is scalar meson exchange. Its main contribution comes
from diagrams with a nucleon pole. This is supposedly contained in the
empirical values of the $c_i$, as discussed in some detail in
Refs.\cite{bkmrev} \cite{ulfmit}. The remaining terms are of the form
depicted in Fig.~6c and include the vertex $\pi S N N$. In the absence
of any empirical indication about the strength of this coupling
constant and the lack of theoretical models thereof, we will set such
terms to zero. It is conceivable that this is a good approximation
since the bulk of scalar meson exchange is expected to be encoded in
some of the terms proportional to the constants $c_i$.

\subsection{The master formula}
\label{sec:master}
We have now assembled  all pieces to give the chiral expansion of the
threshold amplitudes $D_{1,2}$, or equivalently,
${\cal A}_{10}$ and ${\cal A}_{32}$. The chiral corrections to the
amplitudes $D_{1,2}$ follow by combining  
eqs.(\ref{dborn},\ref{d12gtr},\ref{d1l},\ref{d2l},\ref{d1ci},\ref{d2ci},
\ref{d12roper}). We have
\beq
D_i = D_i^{\rm LET} + D_i^{(2)} +{\cal O}(M_\pi^3) 
\, \, , \quad i = 1,2 \, \, ,
\label{dichir}
\eeq
with 
\beq
 D_1^{\rm LET} = \dfrac{g_{\pi N}}{8 m F_\pi^2} \biggl( 1 + \dfrac{7}{2} \mu 
\biggr) \, , \, \, \, \, 
 D_2^{\rm LET} = -\dfrac{g_{\pi N}}{8 m F_\pi^2} \biggl( 3 + \dfrac{17}{2} \mu 
\biggr) \, , 
\label{d12let}
\eeq
and we do not write down again the various pieces contributing to the order
$M_\pi^2$ corrections explicitely. Symbolically, they take the form
\beq
D_i^{(2)} = D_i^{(2,{\rm Born})} + D_i^{(2,{\rm loop})} 
+D_i^{(2, \ell_i)}  + D_i^{(2, c_i)} +D_i^{(2,{\rm GTR})} 
    + D_i^{(2, N^\star)}            \, , \, \, \,
i = 1,2 \, \, ,
\label{d122}
\eeq
with $D_1^{(2,{\rm GTR})} =D_1^{(2, N^\star)} = 0$, 
and we neglect the imaginary parts as
discussed before. The LETs were first derived  in ref.\cite{bkmp}.
With the help of eqs.(\ref{a004},\ref{a024}) as well as eq.(\ref{adef}),
we can make explicit the $\pi \pi$
scattering  lengths (i.e. give the constants $\tilde{d}_I^0$,
cf. eq.(\ref{shift})). This determines the relation between the threshold
$\pi \pi N$ amplitudes and the $\pi \pi$ scattering lengths one order
beyond the improved representation, eq.(\ref{a_Inew}), and reads, 
\begin{equation}
   {\rm Re} \, {\cal A}_{10} 
   = 4 \pi \, \dfrac{g_{\pi N}}{m} \dfrac{a_0^0}{M_\pi^2}
   + \dfrac{37 g_{\pi N} M_\pi}{16 m^2 F_\pi^2} 
   + \Delta_0 \, \dfrac{g_A M_\pi^2}{32 F_\pi^5} 
   + {\cal O}(M_\pi^3) 
\label{master10}
\eeq
\beq
   {\rm Re} \, {\cal A}_{32} =
   - 2 \sqrt{10} \, \pi \, \dfrac{g_{\pi N}}{m} \dfrac{a_0^2}{M_\pi^2}
   + \dfrac{7 \sqrt{10} g_{\pi N} M_\pi}{16 m^2 F_\pi^2} 
   + \Delta_2 \, \dfrac{ \sqrt{10} g_A M_\pi^2}{128 F_\pi^5} 
   + {\cal O}(M_\pi^3) 
\label{master32}
\end{equation}
with 
$$
\Delta_0 = 16 \biggl( 8 \ell_1^r - 4 \ell_2^r 
+ 3 \ell^r_4 \biggr) - \dfrac{F_\pi^2}{m^2} \biggl( \dfrac{31}{2} + 
12 g_A^2 \biggr) + \dfrac{64 F_\pi^2}{m} \biggl(2c_1 - 2c_3 + c_4 +
c_2'' \biggr) +
$$
$$
\dfrac{1}{\pi^2} \biggl[ \bigl( 28 g_A^2 - 16 \bigr)
\ln \dfrac{M_\pi}{\lambda} + \dfrac{41}{6} - \dfrac{49}{3} g_A^2
+ \bigl(14 g_A^2 + \dfrac{1}{4} \bigr) \sqrt{3} \ln (2 + \sqrt{3} )
-32 \, I  \biggr]
$$
\beq
+ 96 \biggl( - (c_1^\star + c_2^\star) \sqrt{R} \dfrac{F_\pi^2}{m^\star - m} +
b_{11} \biggr)
\label{Delta0}
\eeq
$$
\Delta_2 = 32 \biggl( 4 \ell_1^r - 2 \ell_2^r 
 \biggr) - \dfrac{2 F_\pi^2}{m^2} \bigl( 1 + 
12 g_A^2 \bigr) + \dfrac{64 F_\pi^2}{m} \biggl(-4c_1 + 4c_3 + c_4 +
3c_2' + 4c_2'' \biggr) +
$$
\beq
\dfrac{1}{\pi^2} \biggl[ \bigl( 4 g_A^2 - 4 \bigr)
\ln \dfrac{M_\pi}{\lambda} + \dfrac{1}{3} \bigl( 2 g_A^2 -11 \bigr)
+ \bigl(2 g_A^2 - \dfrac{7}{2} \bigr) \sqrt{3} \ln (2 + \sqrt{3} )
+ 40 \, I  \biggr]
\label{Delta2}
\eeq
where the factors in front of $\Delta_{0,2}$ have been chosen such
that the $\Delta_{0,2}$ are numerically of the order ${\cal O}(1)$ and $I$ is
given in eq.(\ref{idef}). For the discussion of the respective
numerical values, we abbreviate the ${\cal O}(M_\pi^2)$ contributions
as follows,
\beq
\Delta_{0,2} = \Delta_{0,2}^{\ell_i} + \Delta_{0,2}^{\rm Born} 
+ \Delta_{0,2}^{c_i}
+ \Delta_{0,2}^{\rm loop} + \Delta_{0,2}^{\rm ct3} \, \, \, ,
\label{Delta}
\eeq
where the various terms can be read off form eqs.(\ref{Delta0},\ref{Delta2}).
Obviously, $\Delta_{2}^{\rm ct3} = 0$.

\section{Results and discussion}
\label{sec:results}
First, we must fix parameters. Throughout, we use $F_\pi = 93$~MeV,
$M_\pi = 139.57$~MeV, $m= 938.27$~MeV, $g_{\pi N} = 13.4$, $g_A =
1.26$ and $m^\star = 1440$~MeV. This leads to $b_{11} = -0.012$.
 
Consider the low--energy constants $c_i$. From the formulae given in
section \ref{sec:lecs} and with the empirical information contained
in table 2.4.7.1 of \cite{lbII}, together with $\sigma_{\pi N}(0) = 45
\pm 9$~MeV \cite{gls} and the scattering volumina 
$c_0  = (0.208 \pm 0.003) \, M_\pi^{-3}$ and 
$d_1 = (-0.069 \pm 0.002) \, M_\pi^{-3}$ 
\cite{lbII}, we arrive  at the numbers given in table~1.
We note that these numbers are typically a factor 1.5 smaller than the
ones form the determination including the loop effects at order
$q^3$ \cite{bkmrev}. The uncertainty in the table reflects the
spread of the various determinations (to order $q^2$)  if possible
(like for $c_1$, $c_3$ and $c_4$), otherwise the 
uncertainty of the empirical input.
\begin{center}  

\renewcommand{\arraystretch}{1.5}

\begin{tabular}{|c|c|c|c|c|} \hline  
 $c_1$ & $c_2'$  & $c_2''$ & $c_3$ & $c_4$   \\ \hline
$-0.64 \pm 0.14$  & $-5.63 \pm 0.10$  & $7.41 \pm 0.10$ 
& $-3.90 \pm 0.09$ & $2.25 \pm 0.09$ \\ 
\hline \end{tabular} 

\medskip

Table 1: The low--energy constants $c_i$. All values in GeV$^{-1}$.

\end{center}
{}From these numbers, we deduce
\beq
\Delta_0^{c_i} = 9.54 \pm 0.21\, \, , \quad
\Delta_2^{c_i} = 1.16 \pm 0.50\, \, .
\label{D02c}
\eeq
Next, we need the low--energy constants from the meson sector, 
$\ell_{1,2,3,4}^r$. We take the $\bar{\ell}_{1,2}$ from the recent
analysis of $K_{\ell 4}$ data beyond one loop \cite{cbg} and the
$\bar{\ell}_{3,4}$ from the classical paper \cite{gl84},
$$
10^3 \, {\ell_1^r} (1 {\rm GeV}) = -5.95 \pm 1.06\, , \quad
10^3 \, {\ell_2^r} (1 {\rm GeV}) =  4.35 \pm 2.75\, , \quad
$$
\beq
10^3 \, {\ell_3^r} (1 {\rm GeV}) =  1.64 \pm 3.80\, , \quad
10^3 \, {\ell_4^r} (1 {\rm GeV}) =  2.29 \pm 5.70\, ,
\label{valli}
\eeq
leading to
\beq
\Delta_0^{l_i} = -0.93 \pm 0.40 \, \, , \quad
\Delta_2^{l_i} = -1.04 \pm 0.22 \, \, .
\label{D02li}
\eeq
at the scale $\lambda = 1$~GeV. If one lets $\lambda$ run from 0.5 to
1.5 GeV, one finds $\Delta_0^{l_i} = -0.49 \ldots -1.05$ and
$\Delta_2^{l_i} = -1.06 \ldots -0.98$. To acount for this, we scale up
the uncertainty of $\Delta_0^{l_i}$ by a factor 1.5. The Born
contribution to $\Delta_{0,2}$ is small, we find
\beq
\Delta_0^{\rm Born} = -0.34 \, \, , \quad
\Delta_2^{\rm Born} = -0.39 \, \, .
\label{D02b}
\eeq
The corresponding loop contributions are also readily evaluated,
\beq
\Delta_0^{\rm loop} = -4.48 \pm 1.58 \, \, , \quad
\Delta_2^{\rm loop} =  1.81 \pm 0.16 \, \, .
\label{D02l}
\eeq
for $\lambda = 1$~GeV as the central value and the uncertainty 
accounts for the
variation if $\lambda$ varies from 0.5 to 1.5 GeV. To end this part, we give
the GTR and Roper contributions,
\beq
\Delta_0^{\rm ct3} = \Delta_0^{\rm GTR} + \Delta_0^{N^\star}
= 0.22 \pm 3.58 \, \, ,
\label{del2ct}
\eeq
where the large uncertainty stems from the Roper couplings  
$c_1^\star + c_2^\star$. From these numbers we can already conclude that the
$1/m$ corrections to the $\pi N$ scattering amplitude 
($\sim \Delta_{0,2}^{c_i}$) play a dominant
role in the total correction of order $M_\pi^2$
together with the Roper excitation. Also, if one
separates out the $\pi \pi$ interactions, their contribution is
comparably small. This indicates that a very accurate determination of
the S--wave $\pi \pi$ scattering lengths will be very difficult (see
below).

Consider now the chiral expansion of the $D_{1,2}$. The LET values
based on eqs.(\ref{d12let}) are $D_1^{\rm LET} = 2.41$~fm$^3$  and
$D_2^{\rm LET} = -6.76$~fm$^3$ compared to 
the empirical ones \cite{bkmrev},
\beq 
D_1^{\exp} = 2.26 \pm 0.12 \, \, {\rm fm}^3 \, \, \quad 
D_2^{\exp} = -9.05 \pm 0.36 \, \, {\rm fm}^3 \, \, .
\label{d12fit}
\eeq
These numbers result from a best fit to the near threshold cross section
data for $\pi^+ p \to \pi^+ \pi^+ n$ and $\pi^- p \to \pi^0 \pi^0 n$
\cite{kernel} \cite{lowe} 
using the correct flux and three--body phase space factors. 
The LET predictions  show the expected pattern of deviation 
from the empirical values (compare the discussion
in Refs.\cite{bkmp} \cite{bkmrev}). The various 
${\cal O}(M_\pi^2)$ corrections to $D_{1,2}$ are summarized in table 2.
\begin{center}  

\renewcommand{\arraystretch}{1.5}

\begin{tabular}{|c|c|c|c|c|c|c|} \hline  
$D^{(2)}_i$ & Born & Loop  & $\ell_i$ & $c_i$ & GTR &  $N^\star$   \\ \hline
$1$ & --0.08 & $0.25 \pm 0.13$  & $-0.17 \pm 0.18$  & $0.24 \pm 0.10$ 
& 0 & 0 \\
$2$ & 0.15 & $0.09 \pm 0.48$  & $0.39 \pm 0.21$  & $-2.85 \pm 0.06$ 
& $0.32$ & $-0.40 \pm 0.90$ \\ 
\hline \end{tabular} 

\medskip

Table 2: Various $M_\pi^2$ corrections to $D_{1,2}$. All values in fm$^3$.

\end{center}
Adding the uncertainties in quadrature, the  predictions for $D_{1,2}$
to order $M_\pi^2$ are
\beq 
D_1^{\rm thy} =  2.65 \pm 0.24 \, \, {\rm fm}^3 \, \, \quad 
D_2^{\rm thy} = -9.06 \pm 1.05 \, \, {\rm fm}^3 \, \, .
\label{d12thy}
\eeq
These results for $D_{1,2}$ are compatible with the empirical values, 
eq.(\ref{d12fit}), within one standard deviation. 
 We notice that there are large cancellations in the
$M_\pi^2$ contributions to $D_1$, whereas the chiral corrections to
$D_2$ at this order are clearly dominated by the terms proportional to
the low--energy constants $c_i$ and the Roper excitation. 
In particular, one reads from table 2 that the loop contribution
to the $I=0$ amplitude ${\cal A}_{10}$ is rather small, roughly -4\%
of the LET value. In contrast to this, the loop corrections for the
$I=0$ $\pi \pi$ scattering length $a_0^0$ are sizeable, about 25\%
of the current algebra value (at $\lambda =1$ GeV). This signals that
contrary to expectations, the reaction $\pi N \to \pi \pi N$ at threshold
is not very sensitive to the $\pi \pi$ final state interactions. Also, if we
disentangle the terms of order $M_\pi^{(n)}$ ($n=0,1,2$) we find that
the convergence of the expansion for $D_1$ is good whereas in $D_2$
one still finds sizeable corrections at $n=2$,
\beq
D_1 = 1.59 \cdot (1. + 0.52 + 0.15) \, \, {\rm fm}^3 \, \, , 
\quad D_2 = -4.76 \cdot (1. + 0.42 + 0.48) \, \, {\rm fm}^3 \, \, , 
\label{d12conv}
\eeq
for the mean values of $D_{1,2}$.
This shows that the chiral expansion for $D_1$ is converging. Matters
are different for $D_2$. One has at least to calculate the
terms of order $M_\pi^3$ before one can draw a clear conclusion about
the accuracy with which $D_2$ can be calculated. This, however,
goes beyond the scope of the present paper. 

\medskip

We now turn to the determination of the S--wave $\pi \pi$ scattering
lengths.\footnote{We add the uncertainty from the theoretical
  determination and the one from the fit to the data in quadrature.}
  Clearly, due to the large $M_\pi^2$ corrections and 
uncertainties (due to the Roper) in $D_2$,
deducing $a_0^0$ from the master formula can only give an indicative
result. We find
\beq
a_0^0 = 0.21 \pm 0.07 \, \, , 
\label{pipia1}
\eeq
to be compared with the current algebra value of $0.16$ and the CHPT
prediction of $0.20 \pm 0.01$\cite{gl84}. We notice that the
theoretical prediction has a much smaller uncertainty than the number
extracted from the $\pi \pi N$ threshold amplitude. 

Matters are different for $a_0^2$ since this quantity is entirely
sensitive to $D_1$.
Adding the empirical and theoretical uncertainties
in quadrature, we find
\beq
a_0^2 = -0.031 \pm 0.007 \, \, , 
\label{pipia}
\eeq
consistent (within one standard deviation) with the one--loop CHPT prediction
of Gasser and Leutwyler,  
$a_0^2 = -0.042 \pm 0.008$ \cite{gl83} \cite{gl84}. 
We remark that for the combination $2a_0^0 - 5a_0^2$ we have $0.577
\pm 0.144$, consistent with the universal curve, $(2a_0^0 -
5a_0^2)_{\rm univ. curve} = 0.614 \pm 0.028$.
This clearly indicates that previously determined
scattering lengths based on the Olsson--Turner model
\cite{pocanic} \cite{burkhard} should not be trusted. 
The lesson to be learned
here is that even with an improved $q^4$ calculation there will remain
sizeable theoretical uncertainties which will make a more
accurate determination of the isospin zero S--wave $\pi \pi$
scattering length
from the $\pi \pi N$ threshold amplitudes  very difficult. In
contrast, one can hope to sharpen the determination of 
$a_0^2$, eq.(\ref{pipia}).

\section{Summary}
\label{sec:summary}
In this paper, we have considered the reaction $\pi N \to \pi \pi N$
at threshold in the framework of heavy baryon chiral perturbation
theory. The pertinent results of this investigation can be summarized
as follows:

\begin{itemize}

\item We have calculated the chiral expansion of the
  threshold amplitudes $D_1$ and $D_2$ (or, equivalently, to ${\cal
    A}_{10}$ and ${\cal A}_{32}$) up--to--and--including  the
  quadratic order in the pion mass.
  This amounts to the first corrections to the low--energy
  theorems derived in Ref.\cite{bkmp}. 
  The resulting values for
  $D_{1,2}$ agree with the empirical ones within one standard deviation.
  For $D_1$, the ${\cal O}(M_\pi^2)$ corrections are 
  small, the corresponding corrections to $D_2$ are sizeable.
  The latter are mostly related to chiral corrections to the $\pi N$
  amplitude and the excitation of the $N^\star (1440)$ resonance.

\item Based on this improved representation for the threshold $\pi \pi
  N$ amplitudes, one can deduce the isospin two S--wave $\pi \pi $ scattering
  length, $a_0^2 = -0.031 \pm
  0.007$. This number is compatible with the one--loop chiral
  perturbation theory results. However, the ensuing uncertainty is
  still sizeable and it will be difficult to further improve upon it.
  Due to the large $M_\pi^2$ corrections in $D_2$, one can only deduce
  a broad range of  values for $a_0^0$ from this calculation,
  $a_0^0 = 0.21 \pm 0.07$. The point here is that
  the threshold $\pi \pi N$ amplitudes are much less sensitive to the
  four--pion vertex than to other effects like resonance excitations
  and uncertainties in the $\pi N$ amplitudes. At present, it appears
  that the threshold $\pi \pi N$ data are best suited to pin down
  the isospin two, S--wave $\pi \pi$ scattering length.

\item As a by--product, we have found a new coupling for the $N^\star
  (1440)$ decay into the nucleon and two pions in the S--wave. It
  differs from the conventionally used one \cite{oset} through its
  explicit energy--dependence (i.e. factors of the pion energies), see
  appendix \ref{appb}. It would be important to disentangle these two 
  couplings to reduce the uncertainty related to the Roper excitation.
  This could eventually be done by photo--exciting the Roper and study
  the decay $N^\star \to N (\pi \pi)_S$ in the threshold region (i.e.
  for $\sqrt{s} \ge m^\star$).

\end{itemize}

Finally, we stress that an order $q^4$ calculation should be performed
to further tighten the chiral predictions and to get a better handle
on the chiral expansion of $D_2$. Also, a consistent calculation to
treat all isospin violating effects (like e.g. $m_d-m_u$, virtual
photons and alike) has to be performed.

\vspace{3cm}

\section*{Acknowledgements}

We are grateful to J\"urg Gasser, Gerhard H\"ohler
and Eulogio Oset for helpful comments.
We are particularly thankful to Marc Knecht, Bashir Moussallam and Jan
Stern for pointing out an inconsistency in the original version of the
manuscript. We thank the Institute for Nuclear Theory at Seattle and the 
Svedberg Laboratory at Uppsala for hospitality extended during 
the completion of this work.

\newpage

\appendix
\def\theequation{\Alph{section}.\arabic{equation}}
\setcounter{equation}{0}
\section{Loop functions}

Here, we present explicit formulae for those loop functions entering the
calculation of $\pi N \to \pi\pi N$ at threshold which are not given in the
review \cite{bkmrev}. 
All propagators are understood to have an infinitesimal negative 
imaginary part in the denominator $(-i\epsilon$). We use dimensional
regularization to compute divergent loop integrals and expand them around $d=4$
space-time dimensions. In the reaction $\pi N \to \pi\pi N$ at threshold we 
encounter the following loop integrals involving three and four propagators, 

\beq {1\over i} \int {d^dl \over (2 \pi)^d} {1\,,\, l_\mu l_\nu \over v\cdot
l\, (M_\pi^2 - l^2) \,(M_\pi^2 -(l+2 q)^2) } = \Phi_0(\omega)\, ,\,g_{\mu\nu}\,
\Phi_3(\omega) + \dots \label{app1} \eeq
\beq {1\over i} \int {d^dl \over (2 \pi)^d} {1\,,\, l_\mu l_\nu \over v\cdot
l\, v\cdot (l+2q)\,(M_\pi^2 - l^2)\,(M_\pi^2 -(l+2 q)^2) } = \Psi_0(\omega)\,
,\, g_{\mu\nu}\, \Psi_3(\omega) + \dots \label{app2} \eeq
with $\omega = v\cdot q$ , $q^2= M_\pi^2 $ and the ellipsis stands for terms
proportional to $v_\mu v_\nu,\, q_\mu q_\nu,\dots$ which are not needed in the
actual calculation. For our purpose the loop functions $\Phi_{0,3}(\omega)$ 
defined above are to be evaluated at $\omega = \pm M_\pi + i0$,
\beq \Phi_0(M_\pi) = {1\over 16 \pi^2 M_\pi} \biggl[ {\pi \over 2} -
\sqrt{3} \ln (2 + \sqrt{3} ) + i \, \pi \sqrt{3} \biggr] \label{app3} \eeq
\beq \Phi_0(-M_\pi) = {1\over 16 \pi^2 M_\pi} \biggl[ {\pi \over 2} +
\sqrt{3} \ln (2 + \sqrt{3} ) \biggr] \label{app3a} \eeq

\beq \Phi_3(M_\pi) = 2 M_\pi \,L+{ M_\pi\over 16 \pi^2} \biggl[2 \ln {M_\pi
\over \lambda} + {\pi \over 6}-{5\over3}  + \sqrt{3} \ln (2 + \sqrt{3} )- i\,
\pi \sqrt3 \biggr] \label{app4} \eeq 
\beq \Phi_3(-M_\pi) = -2M_\pi\, L + {M_\pi\over 16 \pi^2 } \biggl[ -2 \ln
{M_\pi \over \lambda} + {\pi \over 6}+{5\over 3}  - \sqrt{3}
\ln (2 + \sqrt{3} ) \biggr] \label{app5} \eeq
with $L$ defined in eq.(\ref{defL}).
In order to get the values of the functions $\Psi_{0,3}(\omega)$ at $\omega =
M_\pi + i0$ one simply uses the identity
\beq {1\over v\cdot (l+2q) \, v\cdot l} = {1\over 2v\cdot q} \biggl( {1\over
v\cdot l } - {1\over v\cdot (l+2q) } \biggr) \label{app7} \eeq
shifts the loop momentum and finds

\beq \Psi_0(M_\pi) ={1\over 2M_\pi}[ \Phi_0(M_\pi ) - \Phi_0(-M_\pi)] =
{\sqrt3\over 32 \pi^2 M_\pi^2} \bigl[ i \, \pi  - 2  \ln (2 + \sqrt3) \bigr]
\label{app8} \eeq
\beq \Psi_3(M_\pi) = {1\over 2 M_\pi} [ \Phi_3(M_\pi) - \Phi_3(-M_\pi) ] = 2 L
+ {1\over 16 \pi^2} \biggl[2 \ln {M_\pi \over \lambda} -  {5 \over 3} +
\sqrt{3} \ln (2 + \sqrt{3} ) - i\, {\pi \over 2} \sqrt3\biggr] \label{app9}
\eeq 

\vfill \eject

\newpage

\setcounter{equation}{0}
\section{The decay \boldmath$N^\star (1440) \to N (\pi \pi)_S$}
\label{appb}

In this appendix, we discuss in detail the Roper decay into the nucleon and 
two pions in the S--wave since that is not treated in generality in the
present literature. In fact, to order $q^2$,  the pertinent Lagrangian
${\cal L}_{N^\star N \pi \pi}$ contains (at least) two terms,
\beq
{\cal L}_{N^\star N \pi \pi} = c_1^\star \bar{\Psi}_{N^\star} \, \chi_+ \,
\Psi_N  
-  \dfrac{c_2^\star}{{m^\star}^2} (D_\mu D_\nu \bar{\Psi}_{N^\star} )
u^\mu u^\nu \Psi_N +{\rm h.c.} 
\, \,  ,
\label{rop2}
\eeq
with $\Psi_{N,N^\star}$ the relativistic spin--1/2 fields. 
In fact, the second term
is not unique, but for the non--relativistic formulation, 
all relativistically inequivalent forms lead to the same operator
proportional to $(v \cdot u)^2$.
Using
\beq
\chi_+ = M_\pi^2 \biggl( 2 - \dfrac{ \vec{\pi \,}^2}{F_\pi^2} + \ldots
\biggr) \, \, , \quad 
u_\mu = - \dfrac{1}{F_\pi} \, \vec{\tau \,} \cdot \partial_\mu \, \vec{\pi \,}
+ \ldots \, \, ,  
\eeq
one finds that the first (commonly used) coupling is
energy--independent whereas the second depends on $\omega_1 \,
\omega_2$, the product of the energies of the two pions. Denoting by
$\Gamma_9$ the partial width $\Gamma (N^\star \to N (\pi \pi)_S)$, we
find using eq.(\ref{rop2}),
\beq
\Gamma_9 = \dfrac{3}{16 \pi^3 F_\pi^4} \int \int_{z^2 < 1}  d\omega_1
  d\omega_2 \, ( m^\star + m - \omega_1 -\omega_2) \bigl( c_1^\star 
\, M_\pi^2 + c_2^\star \, \omega_1 \omega_2 \bigr)^2 \, \, \, ,
\label{ps}
\eeq
with
\beq
z = \dfrac{\omega_1 \omega_2 - m^\star(\omega_1+\omega_2) + M_\pi^2
  + ({m^\star}^2 -m^2)/2}{\sqrt{(\omega_1^2- M_\pi^2) (\omega_2^2- M_\pi^2)}}
\, \, . \eeq
Performing the integrals in eq.(\ref{ps}), we have
\beq
\Gamma_9 = \biggl[ 0.498 {c_1^\star}^2 + 2.708 {c_1^\star} {c_2^\star}
+ 3.714 {c_2^\star}^2 \biggr] \, 10^{-3} \, \, {\rm GeV}^3 \, \, ,
\label{rn2pi}
\eeq
which is somewhat surprising since one would not expect the factor
of pion energies leading to such a difference compared
 to the energy--independent
coupling. Eq.(\ref{rn2pi}) defines a rather elongated ellipse.
The determination from  the width does not fix the signs, we choose it
to be negative for $c_1^\star$ since the first term in the
Lagrangian eq.(\ref{rop2}) can be considered as 
"transition $\sigma$--term" (analogous to the $c_1$ term in the dimension two
pion--nucleon Lagrangian) and we know that $c_1 < 0$. Similarly, we
set $c_2^\star >0$. For the threshold amplitude $D_2$, only the sum of
these two coupling constants is relevant. In the absence of any
further empirical information, we take a mean value between the two
extrema, $c_1^\star = 0$ or $c_2^\star =0$, and allow for a large
uncertainty to accomodate both possibilities. This leads to
\beq
c_1^\star + c_2^\star = -1.56 \pm 3.35 \, \, {\rm GeV}^{-1} \, \, \, ,
\label{c12rop}
\eeq
which we use in the main text. These two couplings could be
disentangled by a precise study of the energy dependence of the Roper
decay in the threshold region.

\setcounter{equation}{0}
\section{Calculation of \boldmath$D_{1,2}^{c_i}$ 
in the heavy mass formulation}
\label{appc}

In this appendix, we briefly sketch the derivation of the contribution 
$D_{1,2}^{c_i}$, eqs.(\ref{d1ci},\ref{d2ci}),
 within the framework of the heavy nucleon Lagrangian.
The starting point is the effective heavy nucleon action to order $q^3$
(for details, see appendix A of Ref.\cite{bkkm}),
\beq
S_{\pi N} = \int d^4 x \, \bar{N} \biggl[ {\cal A}^{(1)} + {\cal A}^{(2)}
+ {\cal A}^{(3)} + \dfrac{1}{2m} \bigl( \gamma_0 {{\cal B}^{(1)}}^\dagger
\gamma_0 {\cal B}^{(2)} + {\rm h.c.} \bigr)  + \ldots \biggr] \, N \, \, ,
\eeq
where the ellipsis stands for terms not contributing to the $c_i / m$ 
corrections. The ${\cal A}^{(i)}$ and ${\cal B}^{(i)}$ are quantities
of order $q^i$. While the ${\cal A}^{(i)}$ connect the ``large''
components of the heavy nucleon fields, the ${\cal B}^{(i)}$ give the
transitions form the ``large'' to the ``small'' components before one
integrates out the latter (for details, see ref.\cite{bkkm}).
 There are three different structures contributing  to 
$D_{1,2}^{c_i}$ (from now on, we drop the superscript '$c_i$'). 
First, there are terms of the type
\beq
{\cal A}^{(2)} \frac{1}{v \cdot \ell} \frac{1}{2m} 
{{\cal B}^{(1)}}^\dagger \gamma^0 {\cal B}^{(1)} \, \, ,
\eeq
which lead to
\beq
D_1 =  2 \, P \, (-c_1 + c_2' + c_2'' + c_3 ) \, \, ,
\eeq
with 
\beq
P = \dfrac{g_A M_\pi^2 }{ m F_\pi^3} \quad .
\eeq
Second, the combination
\beq
\dfrac{1}{2m} \gamma^0 {{\cal B}^{(1)}}^\dagger \gamma^0
{\cal B}^{(2)} + {\rm h.c.} \, \, \, ,
\eeq
contributes to $D_1$ and $D_2$,
\beq
D_1 = \frac{1}{2} P \, (-c_2' + c_4)  \, \, \quad D_2 = -P \,  c_4 \, \, ,
\eeq
and third, there is a $D_2$ contribution from
\beq
{\cal A}^{(3)} \frac{1}{v \cdot \ell} {\cal A}^{(1)} \, \, ,
\eeq
which reads
\beq
D_2 = -P \, (c_2' + 2 c_2'' ) \quad.
\eeq
Putting pieces together, we arrive at eqs.(\ref{d1ci},\ref{d2ci}).

\newpage

\noindent{\Large {\bf Figure Captions}}

\bigskip

\begin{enumerate}

\item[Fig.1] Born graphs. Solid and dashed lines denote nucleons and
pions, respectively. The upper two diagrams are the contact and pion--pole
graphs. The others are suppressed by powers of $1/m$ in HBCHPT.

\item[Fig.2] Renormalization of $G_\pi$, $F_\pi$ and $M_\pi$ to one
loop. Crosses denote counter term insertions. The double and wiggly lines
represent the pseudoscalar density and the axial current, in order. For 
other notations, see Fig.~1.

\item[Fig.3] Renormalization of the pion--nucleon vertex.
Counter term insertions are not shown. For notations, see Fig.~1.

\item[Fig.4] One--loop diagrams non--vanishing at threshold. Graphs
with the two out--going pion lines interchanged are not shown. 
For notations, see Fig.~1.

\item[Fig.5] $1/m$ suppressed contributions from the 
relativistic dimension two
pion--nucleon Lagrangian. The circle--cross denotes an insertion from
$ {\cal L}_{\pi N}^{(2)}$ proprtional to one of the low--energy constants
$c_i$. For notations, see Fig.~1.

\item[Fig.6] Resonance saturation. (a) and (b) refer to nucleon excitations,
like the $\Delta (1232)$ or the $N^\star (1440)$ (double lines) and (c)
to a t--channel meson exchange (double line denoted 'R').

\end{enumerate}
 
\newpage
$\;$

\vskip 3cm

\begin{figure}[bht]
\centerline{
\epsffile{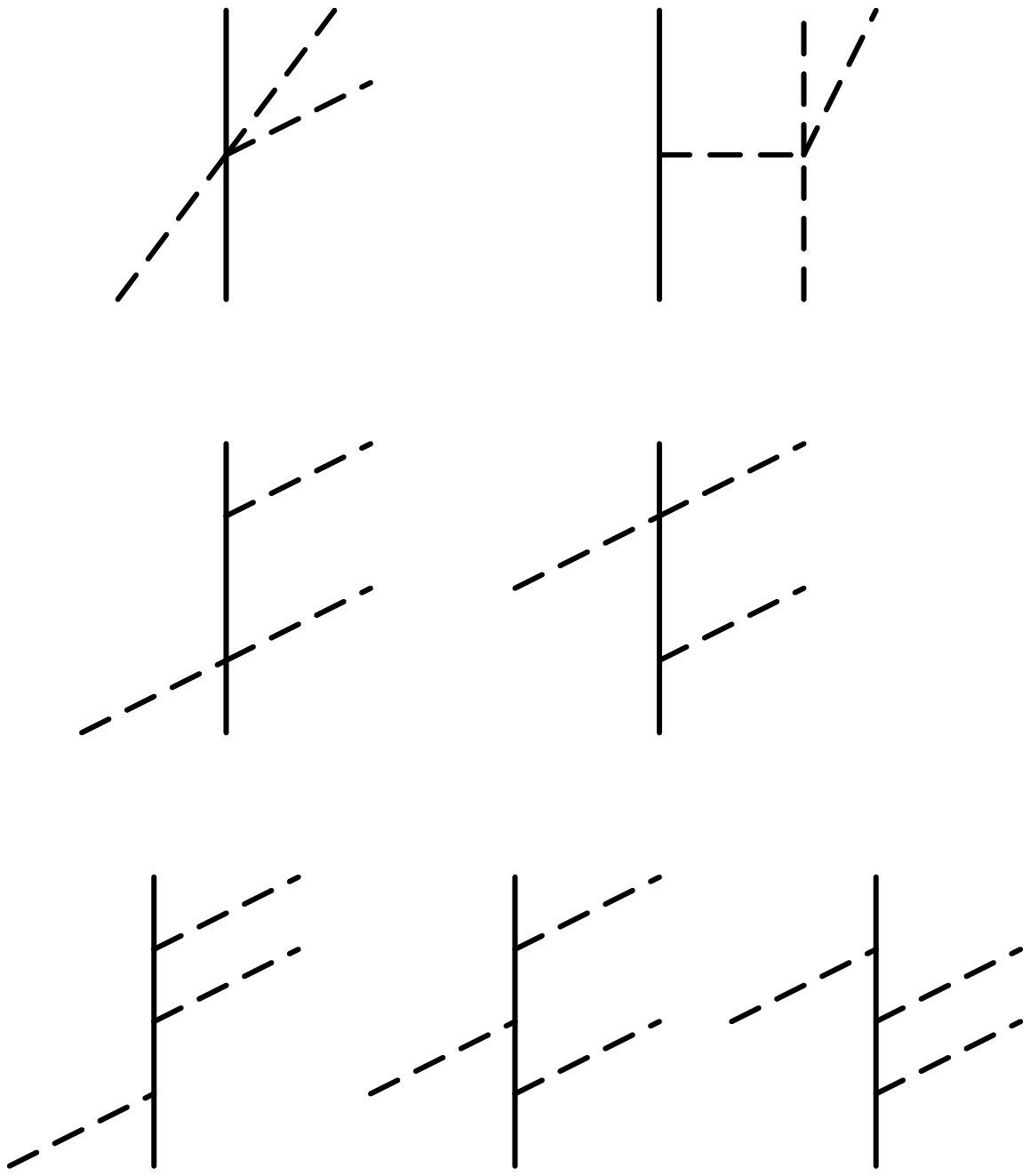}
}
\vskip 2.5cm

\centerline{\Large Figure 1}
\end{figure}
 

\newpage

$\,$

\vskip 2cm

\begin{figure}[bht]
\centerline{
%
%
\epsfysize=5in
\epsffile{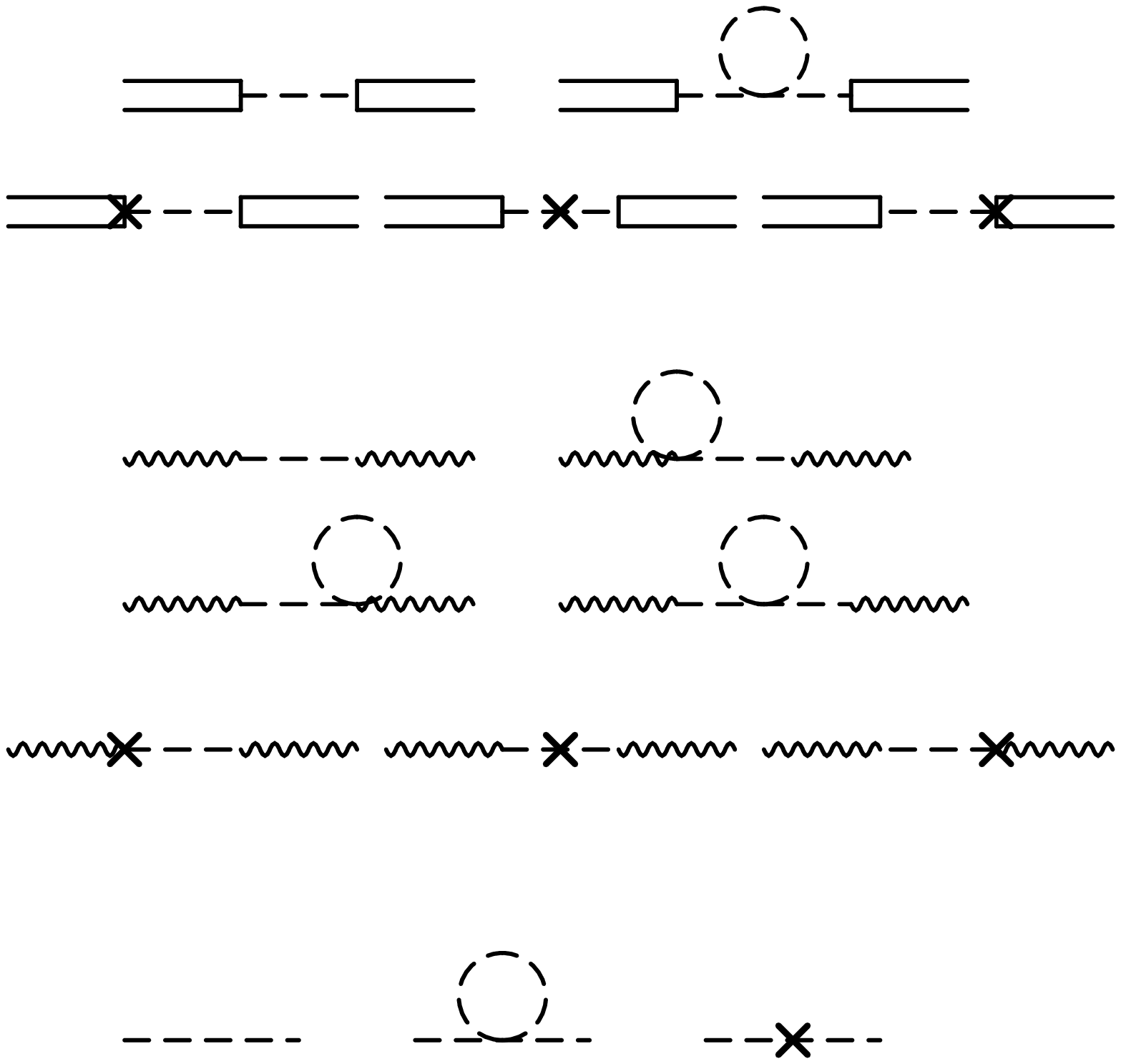}
}
\vskip 2.5cm

\centerline{\Large Figure 2}
\end{figure}
 


\newpage

$\;$\vspace{2cm}

\begin{figure}[bht]
\centerline{
\epsfysize=4.5in
\epsffile{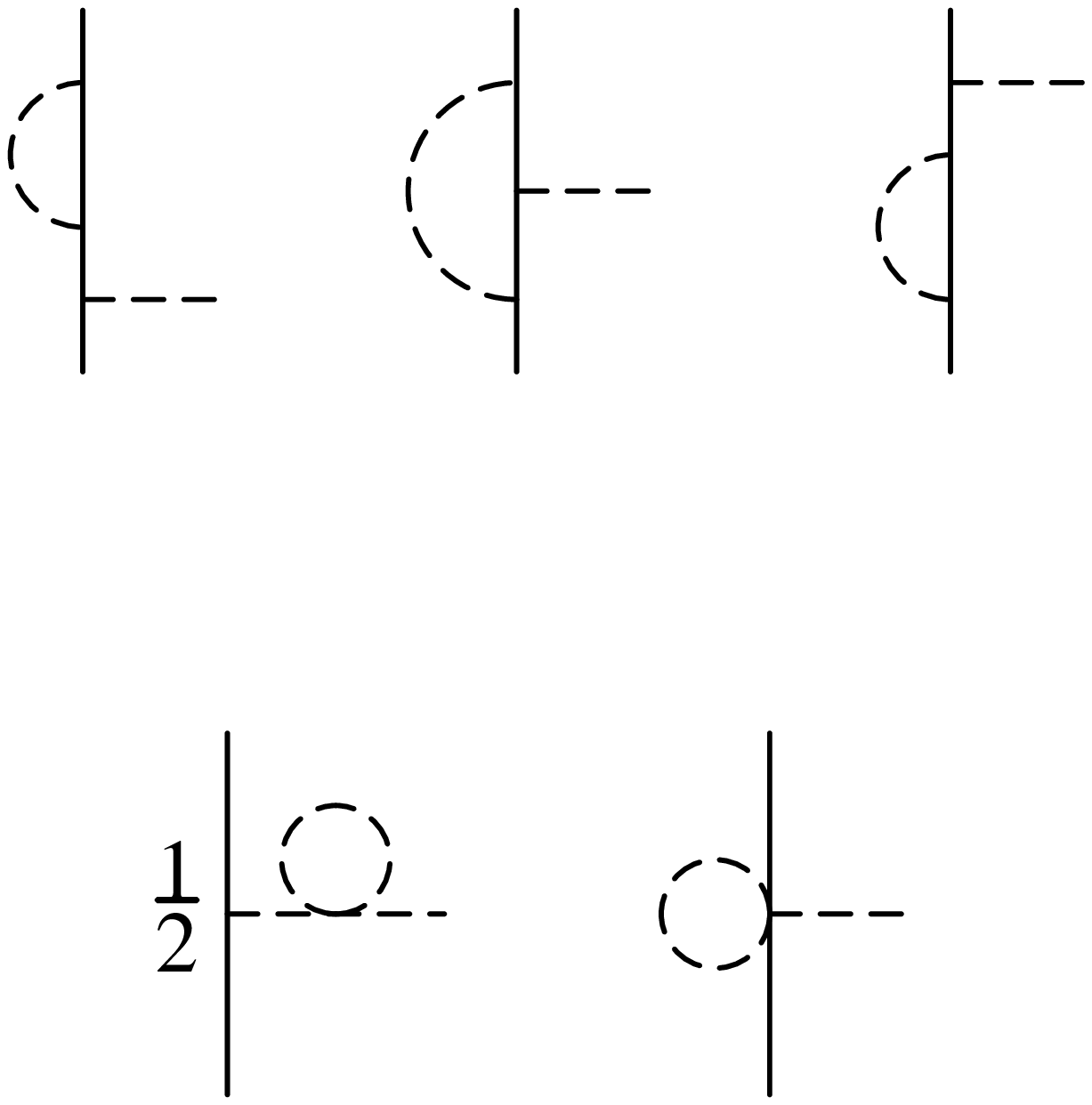}
}
\vskip 2.5cm

\centerline{\Large Figure 3}
\end{figure}
 

\newpage
$\;$ \vskip 2cm

\hskip 3in
%
\epsfxsize=2.5in
\epsfysize=3in
\epsffile{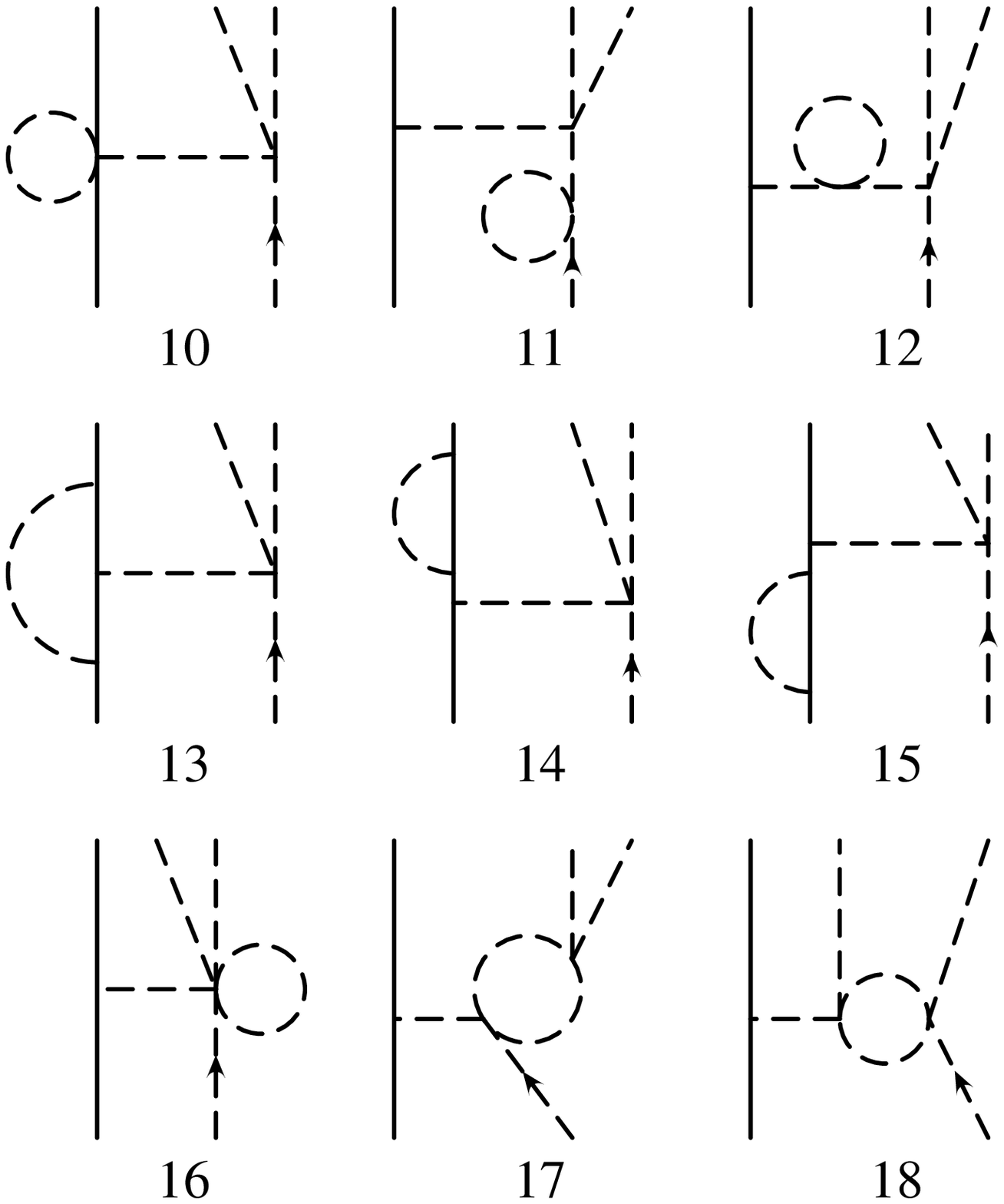}

\vspace{-3in}

\epsfxsize=2.5in
\epsfysize=3in
\epsffile{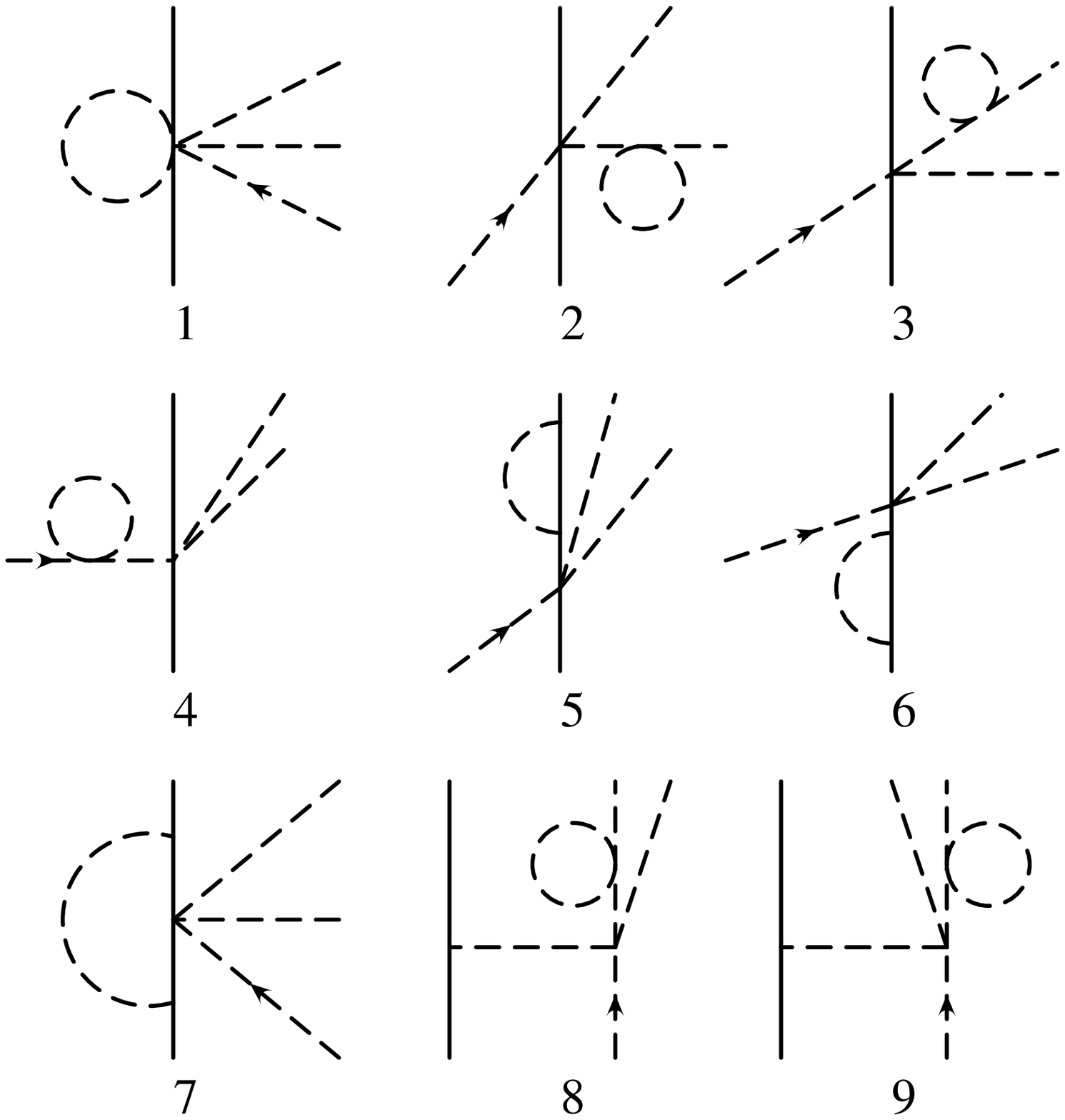}
\vskip 0.2cm

\hskip 3in
%
\epsfxsize=2.5in
\epsfysize=3in
\epsffile{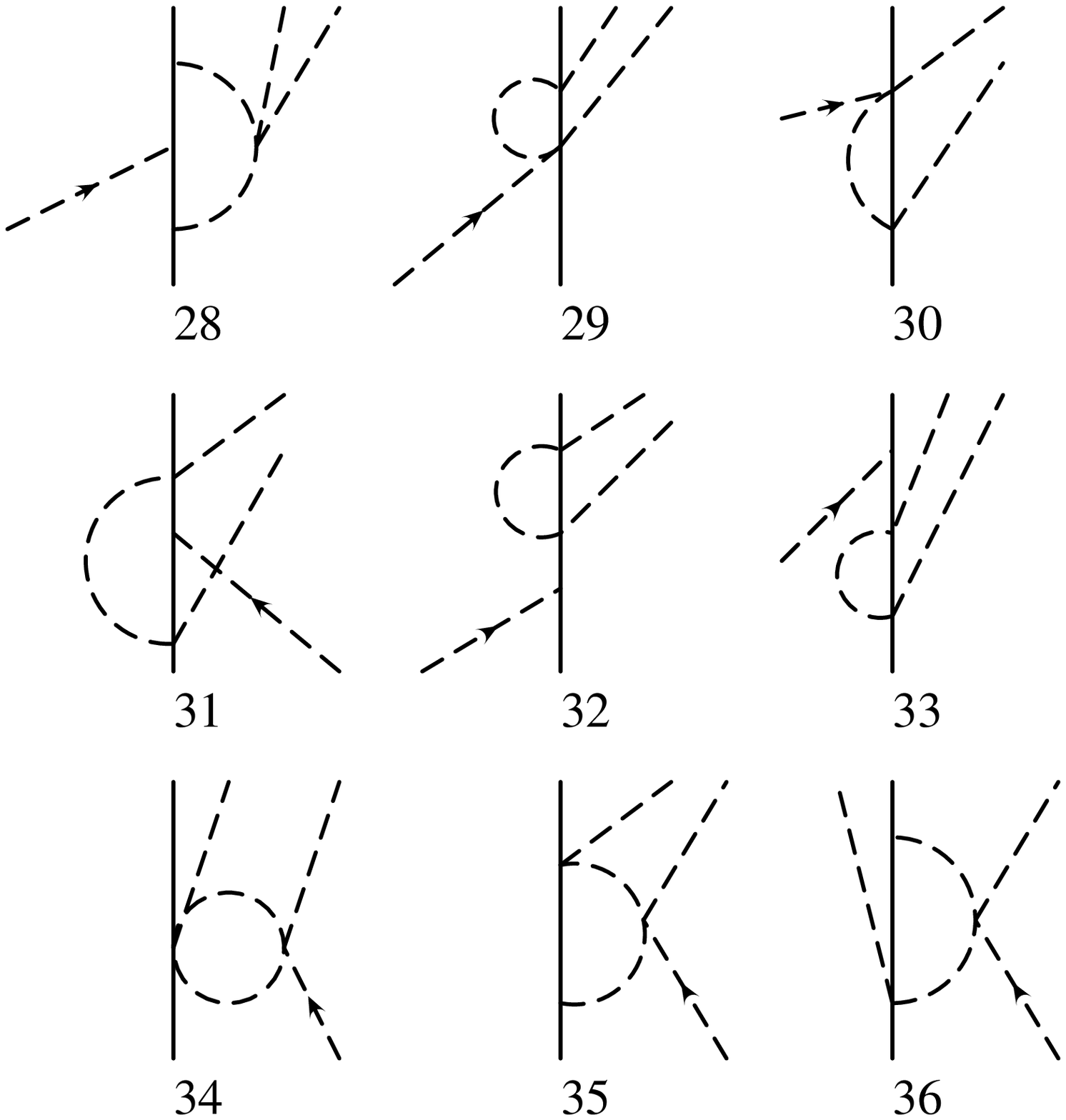}

\vspace{-3in}

\epsfxsize=2.5in
\epsfysize=3in
\epsffile{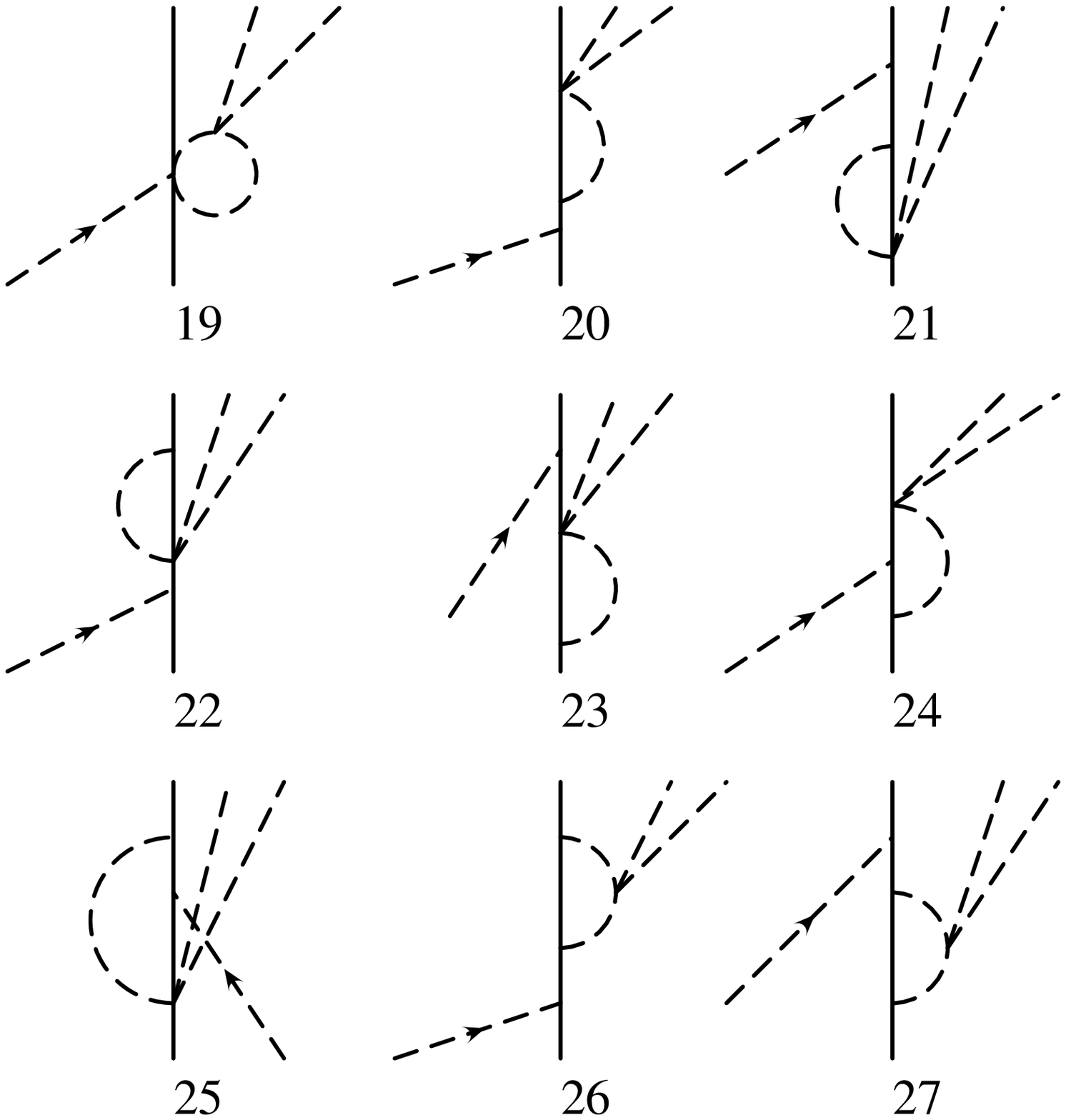}
\vskip 2.5cm

\centerline{\Large Figure 4}

\newpage


\hskip 1.5in
\epsfxsize=2.9in
\epsfysize=3in
\epsffile{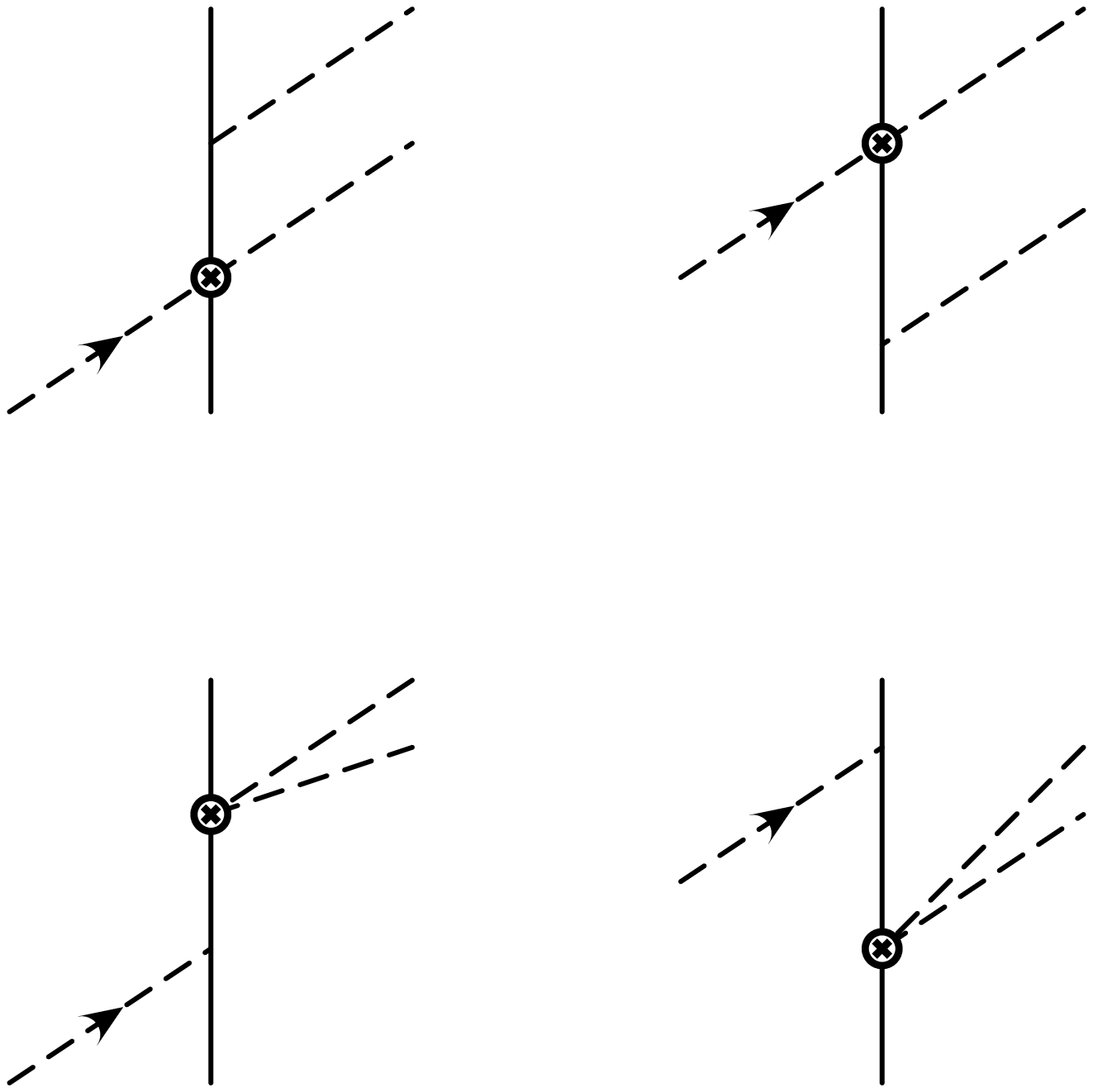}
\vskip 1cm

\centerline{\Large Figure 5}
 


$\;$\vspace{2cm}


\hskip 1in
\epsfxsize=4in
\epsfysize=3in
\epsffile{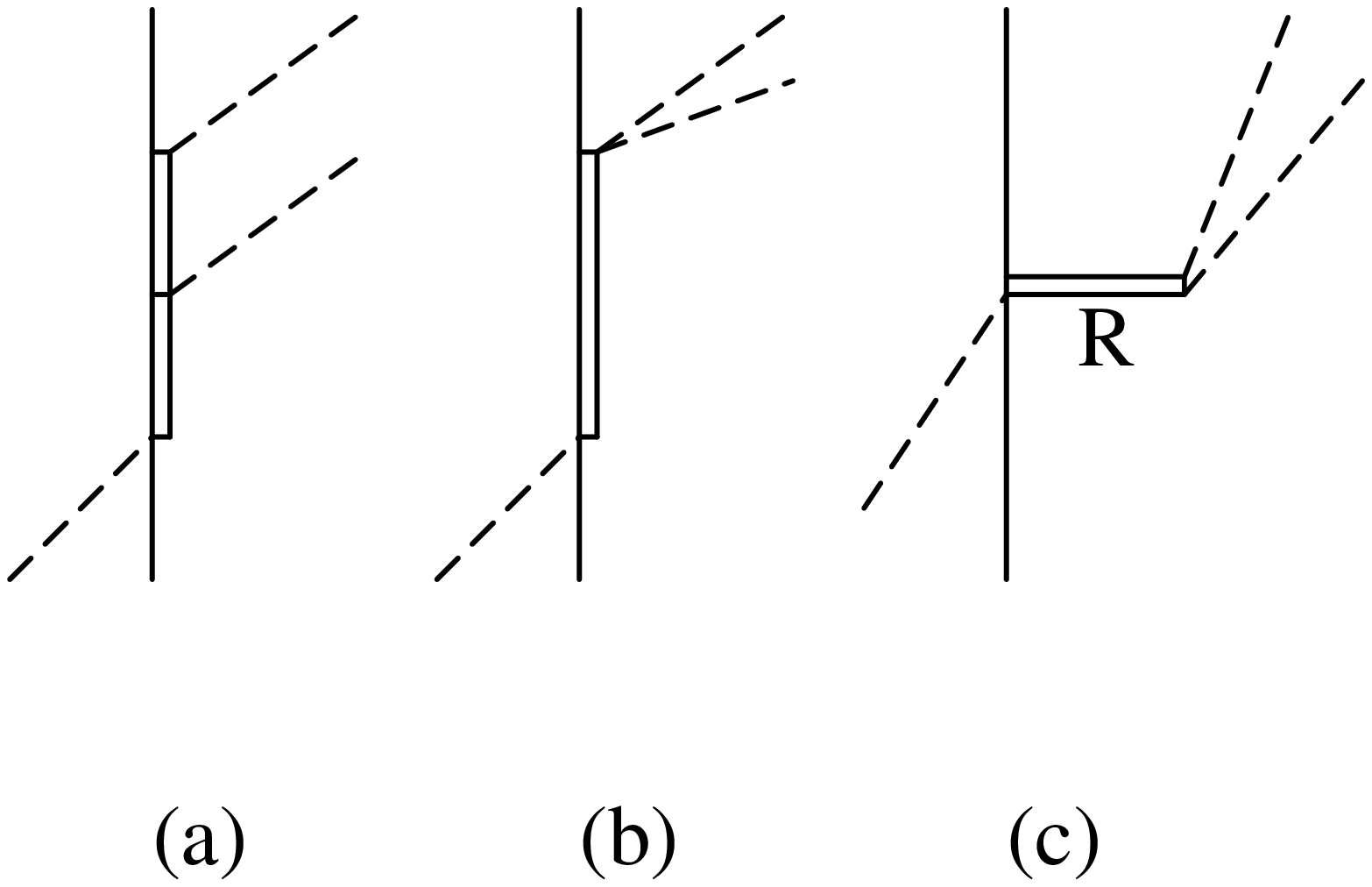}
\vskip 1cm

\centerline{\Large Figure 6}
 


\newpage
\begin{thebibliography}{99}
\frenchspacing

\bibitem{wein1} S.~Weinberg, Phys. Rev. Lett. {\bf 17}, 616 (1966)

\bibitem{gl83}J. Gasser and H. Leutwyler, Phys. Lett. {\bf 125B}, 325 (1983).

\bibitem{gl84}J. Gasser and H. Leutwyler, Ann. Phys. (NY) {\bf 158},
  142 (1984).

\bibitem{stern} J. Stern, H. Szadijan, 
and N. Fuchs, Phys.\ Rev.\ {\bf D38}, 2195 (1988).

\bibitem{orsay}M. Knecht and J. Stern, Orsay preprint IPNO-TH-94-53,
  to be published in the second edition of the DAPHNE physics
  handbook, eds. L. Maiani, G. Pancheri and N.Paver.

\bibitem{ulfrev}Ulf--G. Mei{\ss}ner, Rep. Prog. Phys. {\bf 56}, 903
  (1993).

\bibitem{russ} A.A. Bolokhov, V.V. Vereshagin and S.G. Sherman, 
Nucl. Phys. {\bf A530}, 660 (1991).

\bibitem{wein2} S.~Weinberg, Phys. Rev. Lett. {\bf 18}, 188 (1967);
Phys. Rev. Lett. {\bf 18}, 507 (1967); Phys. Rev. {\bf 166}, 1568 (1968).

\bibitem{ot} M.G. Olsson and Leaf Turner, Phys. Rev. Lett {\bf 20}, 1127
(1968); Phys. Rev. {\bf 181}, 2141 (1969); Phys. Rev. Lett. {\bf 38}, 296
(1977).

\bibitem{obkm} M.G. Olsson, Ulf-G. Mei\ss ner, N. Kaiser and V. Bernard,
"On the interpretation of the $\pi N \to \pi\pi N$ data near
threshold", 
preprint CRN 95-13,
MADPH-95-866 and TK 95 07,  March 1995, to appear in the $\pi N$ Newsletter.

\bibitem{kernel} G. Kernel et al., Z.~Phys. {\bf C48}, 201 (1990);
M.~Sevior et al., Phys. Rev. Lett. {\bf66}, 2569 (1991);
G.~Smitt et al., (CHAOS at TRIUMF, 1994).  $[\pi^+p\to\pi^+\pi^+n]$

\bibitem{pocanic} D.~Po\v cani\'c et al., 
Phys. Rev. Lett. {\bf 72}, 1156 (1993);
G.~Smitt et al., (CHAOS at TRIUMF, 1994).  $[\pi^+p\to\pi^+\pi^0p]$

\bibitem{kernel2} G. Kernel et al., Phys. Lett. {\bf B216}, 244 (1989);
G.~Smitt et al., (CHAOS  at TRIUMF, 1994);
G.~Rebka et al., (LAMPF, 1994). $[\pi^-p\to\pi^-\pi^+n]$

\bibitem{kernel3} G. Kernel et al., Phys. Lett. {\bf B225}, 198 (1989);
G.~Smitt (CHAOS at TRIUMF), 1994). $[\pi^-p\to\pi^-\pi^0p]$

\bibitem{lowe} J. Lowe et al., Phys. Rev. {\bf C44}, 956 (1991).
$[\pi^-p\to\pi^0\pi^0n]$

\bibitem{burkhard} H. Burkhard and J. Lowe, Phys. Rev. Lett. {\bf 67}, 2622
(1991).

\bibitem{bkmp} V. Bernard, N. Kaiser and Ulf-G.~Mei\ss ner, Phys. Lett.
{\bf B332}, 415 (1994); (E) {\bf B338}, 520 (1994).

\bibitem{bkmrev} V. Bernard, N. Kaiser and Ulf-G.~Mei\ss ner,
Int. J. Mod. Phys. {\bf E4}, 193 (1995).

\bibitem{jm} E. Jenkins and A.V. Manohar, Phys. Lett. {\bf B255}, 558 (1991).

\bibitem{bkkm} V. Bernard, N. Kaiser, J. Kambor and Ulf-G. Mei\ss ner,
Nucl. Phys. {\bf B388}, 315 (1992).

\bibitem{watson} K.H.~Watson, Phys. Rev. {\bf 95}, 228 (1954).

\bibitem{eckerrev} G. Ecker, Prog. Nucl. Part. Phys. {\bf 35}, 1
  (1995).

\bibitem{prag} Ulf-G. Mei\ss ner, Czech. J. Phys. {\bf 45}, 153 (1995).

\bibitem{ecker} G. Ecker, Phys. Lett. {\bf B336}, 508 (1994). 

\bibitem{bkms} V. Bernard, N. Kaiser, Ulf-G. Mei\ss ner and A. Schmidt,
Nucl. Phys. {\bf A580}, 475 (1994).

\bibitem{ulfmit} Ulf--G. Mei{\ss}ner, in "Chiral Dynamics: Theory
and Experiment", A. Bernstein and B.R. Holstein (eds.), Springer,
Heidelberg, 1995.

\bibitem{gss} J. Gasser, M.E. Sainio and A. ${\check{\rm S}}$varc, 
Nucl. Phys. {\bf B307}, 779 (1988).

\bibitem{gm} J. Gasser and Ulf-G. Mei{\ss}ner, Phys. Lett. {\bf B258},
  258 (1991).

\bibitem{krause} A. Krause, Helv. Acta Phys. {\bf 63}, 3 (1990).

\bibitem{lbII} G. H\"ohler, in Landolt--B\"ornstein, vol.9b2,
ed. H. Schopper, Springer, Berlin 1983.

\bibitem{tomo} Y. Tomozawa, Nuovo Cim. {\bf 46A}, 707 (1966).

\bibitem{reso} G. Ecker, J. Gasser, A. Pich and E. de Rafael,
Nucl. Phys. {\bf B321}, 311 (1989);\\
J.F. Donoghue, C. Ramirez and G. Valencia,
  Phys. Rev. {\bf D39}, 1947 (1989).

\bibitem{pdg} Particle Data Group, Phys. Rev. {\bf D50}, 1173 (1994).

\bibitem{hoehrop} G. H\"ohler, $\pi N$ Newsletter {\bf 9}, 1 (1993).

\bibitem{oset}E. Oset and M.J. Vicente--Vacas, Nucl. Phys. {\bf A446},
  584 (1985).

\bibitem{oset2}J.A. Gomez Tejedor and E. Oset, Nucl. Phys. {\bf A571},
  667 (1994).

\bibitem{manl} D.M. Manley, R.A. Arndt, Y. Goradia and V.L. Teplitz,
 Phys. Rev. {\bf D30}, 904 (1984).

\bibitem{gls}J. Gasser, H. Leutwyler and M.E. Sainio, Phys. Lett. {\bf B253},
  252 (1991).

\bibitem{cbg}J. Bijnens, G. Colangelo  and J. Gasser, Nucl. Phys {\bf
    B427}, 427 (1994).


\end{thebibliography}
\end{document}